\documentclass[prb,twocolumn,showpacs,amsmath,amssymb]{revtex4}
\usepackage{graphicx}
\begin{document}

\title{
Ultranarrow resonance in Coulomb drag between quantum wires\\ at coinciding densities
}
\author{A. P. Dmitriev$^{1}$}
\author{I. V. Gornyi$^{1,2,3,4}$}
\author{D. G. Polyakov$^{2}$}
\affiliation{
$^{1}$A.F.Ioffe Physico-Technical Institute, 194021 St.Petersburg, Russia\\
$^{2}$Institut f\"ur Nanotechnologie, Karlsruhe Institute of Technology, 76021 Karlsruhe, Germany\\
$^{3}$L.D.Landau Institute for Theoretical Physics, 119334 Moscow, Russia\\
$^{4}$\mbox{Institut f\"ur Theorie der Kondensierten Materie, Karlsruhe Institute of Technology, 76128 Karlsruhe, Germany}
}


\begin{abstract}
We investigate the influence of the chemical potential mismatch $\Delta$ (different electron densities) on Coulomb drag between two parallel ballistic quantum wires. For pair collisions, the drag resistivity $\rho_{\rm D}(\Delta)$ shows a peculiar anomaly at $\Delta=0$ with $\rho_{\rm D}$ being finite at $\Delta=0$ and vanishing at any nonzero $\Delta$. The ``bodyless" resonance in $\rho_{\rm D}(\Delta)$ at zero $\Delta$ is only broadened by processes of multi-particle scattering. We analyze Coulomb drag for finite $\Delta$ in the presence of both two- and three-particle scattering within the kinetic equation framework, focusing on a Fokker-Planck picture of the interaction-induced diffusion in momentum space of the double-wire system. We describe the dependence of $\rho_{\rm D}$ on $\Delta$ for both weak and strong intrawire equilibration due to three-particle scattering.
\end{abstract}

\maketitle

\section{Introduction}
\label{intro}

When two conductors are placed close to each other, so that electron density fluctuations in one of them interact with those in the other, this can change the transport properties of the system even if the conductors do not exchange electrons. In particular, sending electric current through one of the conductors can initiate transport of electrons in the other, mediated solely by frictional forces. Charge transport induced by Coulomb interactions between electrons in different conductors, known as Coulomb drag, was predicted about four decades ago for two-dimensional geometry of two parallel conducting sheets.\cite{pogrebinskii77} First observed in a quantum well in proximity to essentially a three-dimensional conductor\cite{solomon89} and in a double quantum well,\cite{gramila91} Coulomb drag has been extensively studied in electron systems of various geometry, see Ref.~\onlinecite{narozhny15} for a recent review.

Coulomb drag in two macroscopically homogeneous electron systems parallel to each other is parametrized by the drag resistivity $\rho_{\rm D}$ defined conventionally as $\rho_{\rm D}=-E_2/j_1$, where $j_1$ is the current density in (``active") conductor 1 and $E_2$ is the electric field in (``passive") conductor 2 under the condition that the current density therein is zero. With this definition, $E_2$ compensates for the friction force induced in conductor 2 by the flow of electrons in conductor 1. The essence of this phenomenon is captured within the conventional theoretical framework\cite{gramila91,jauho93,zheng93,kamenev95,flensberg95} calculating the friction force at the golden-rule level, perturbatively in the (dynamically screened) interaction between two electron systems. However, it has become clear recently\cite{dmitriev12} that there is a conceptually important limitation imposed on this approach by the rate of thermal equilibration within each of the systems. Although conventional theory for Coulomb drag, providing a Kubo-type formula\cite{jauho93,zheng93,kamenev95,flensberg95} for $\rho_{\rm D}$ (relating $\rho_{\rm D}$ to the dynamical correlations in density fluctuations at equilibrium), may look like having similar status as the fluctuation-dissipation theorem, it tacitly implies that thermal equilibration is a faster process compared to Coulomb drag. In fact, conventional theory hinges on the assumption that electron density fluctuations in the active conductor are equilibrium in the frame moving with the drift velocity.

The inherent relation between thermalization and Coulomb drag was highlighted in Ref.~\onlinecite{dmitriev12} as the essential ingredient of the calculation of $\rho_{\rm D}$ for two parallel quantum wires\cite{flensberg98,nazarov98,gurevich98,gurevich00,klesse00,ponomarenko00,komnik01,trauzettel02,pustilnik03,schlottmann04,fuchs05,fiete06,peguiron07,
aristov07,rozhkov08,dmitriev12} (for experiments on Coulomb drag between quantum wires see Refs.~\onlinecite{debray01,yamamoto06,laroche10,yamamoto12,laroche14}). This geometry provides a striking example of the failure of conventional theory for Coulomb drag. In one dimension, there are only two possibilities: either electron-electron scattering due to interwire interaction changes the chirality of electrons or not. Electron-electron scattering with small-momentum transfer (much smaller than the Fermi momentum $k_F$) is, then, the alternative mechanism of Coulomb drag if interaction-induced backscattering on the Fermi surface can be neglected (which is the situation encountered when the distance between the wires $a$ is sufficiently large: the backscattering amplitude decreases, with increasing $a$, exponentially in the parameter $k_Fa$). Within the conventional approach, interwire interactions without changing the chirality of electrons give a finite contribution\cite{pustilnik03,aristov07,rozhkov08} to $\rho_{\rm D}$. However, as shown in Ref.~\onlinecite{dmitriev12}, slow thermal equilibration in one dimension severely restricts the applicability of the perturbative approach to Coulomb drag; indeed, to the extent that $\rho_{\rm D}$ exactly vanishes for the case of electron-electron interactions preserving the chirality.

Here, we study Coulomb drag between quantum wires with different electron densities. This problem gives another remarkable example of inadequacy of the conventional approach to Coulomb drag. In the conventional formalism, which relates $\rho_{\rm D}$ to dynamical cross-correlations (at the lowest order in interwire interaction) between thermal density fluctuations, the characteristic scale of the Fermi energy difference $\Delta$ between the active and passive conductors in the function $\rho_{\rm D}(\Delta)$ is generically given by the temperature. As will be demonstrated below, Coulomb drag mediated by pair collisions (i.e., at the lowest order in interwire interaction for the collision integral in the kinetic-equation formalism) actually vanishes for arbitrary $\Delta\neq 0$, even though it is finite\cite{dmitriev12} for identical wires. This zero-width resonance in the function $\rho_{\rm D}(\Delta)$ at $\Delta=0$ will be shown to be a peculiar feature of two-particle scattering. Taking three-particle scattering into account broadens the resonance, and if three-particle scattering becomes sufficiently strong, leads to a more conventional behavior of $\rho_{\rm D}(\Delta)$ with a characteristic scale of $|\Delta|\sim T$. In a wide range of the parameters of the problem, $\rho_{\rm D}$ is then determined by the rate of intrawire equilibration\cite{dmitriev12} due to three-particle scattering, whose dependence on $\Delta$ we also calculate below.

The paper is organized as follows. Section~\ref{s2} deals with the dependence of $\rho_{\rm D}$ on $\Delta$ for friction mediated by pair collisions. In Sec.~\ref{s3}, we include three-particle scattering in the collision integral for arbitrary $\Delta$. In Sec.~\ref{s4}, we discuss the zero modes in the problem of Coulomb drag. In Sec.~\ref{s5}, we obtain the matrix structure of the drag resistivity matrix in the dc limit. In Sec.~\ref{s6}, we write the Fokker-Planck kinetic equation for arbitrary $\Delta$ in the presence of both two- and three-particle scattering. In Sec.~\ref{s7}, we solve an exactly solvable model for the Fokker-Planck formulation of the problem. In Sec.~\ref{s8b}, we obtain the general relation between $\rho_{\rm D}$ and the regular, in the dc limit, part of the conductivity matrix. In Sec.~\ref{s8}, we analyze the broadening of the ultranarrow resonance in $\rho_{\rm D}(\Delta)$ at $\Delta=0$ by three-particle scattering. In Sec.~\ref{s9}, we describe the dependence of $\rho_{\rm D}$ on $\Delta$ for the case when intrawire equilibration due to three-particle scattering is strong enough to destroy the anomalously narrow resonance at $\Delta=0$. Our main results are summarized in Sec.~\ref{s10}. Some of the technical details are moved to the Appendixes.

\section{Two-particle scattering}
\label{s2}

Throughout the paper, we assume that the temperature $T$ and the electron dispersion relation $\epsilon=k^2/2m$ in two wires are the same, so that the difference between the wires is parametrized by the difference in their chemical potentials $\Delta=\epsilon_{F1}-\epsilon_{F2}$, where $\epsilon_{F\sigma}$ is the chemical potential in wire $\sigma=1,2$. It is convenient to introduce the function $g_\sigma (k)$ related to the distribution function $f_\sigma(k)$ in wire $\sigma$ by
\begin{equation}
f_\sigma=f_{T\sigma}+g_\sigma T\partial_\epsilon f_{T\sigma}~,
\label{4.1}
\end{equation}
where $f_{T\sigma}=[1+e^{(\epsilon-\epsilon_{F\sigma})/T}]^{-1}$ is the thermal distribution in wire $\sigma$. For the case of two-particle scattering, the kinetic equation for $g_\sigma$ reads:
\begin{equation}
-i\omega g_\sigma-eE_\sigma k/mT={\rm st}^{(2)}_\sigma~,
\label{4.2}
\end{equation}
where $E_\sigma$ is the electric field ($e>0$) in wire $\sigma$ and the (linearized) two-particle collision integral ${\rm st}^{(2)}_\sigma(1)$ at the momentum $k_1$ [with $g_\sigma(1)\equiv g_\sigma (k_1)$, etc.] is given by
\begin{eqnarray}
{\rm st}^{(2)}_\sigma(1)\!\!&=&\!\! {1\over \zeta^2_\sigma(1)}\,\sum_{21'2'}W^{(2)}_\sigma\,\delta_2(\ldots)\,\nonumber\\
&\times&\!\![\,g_\sigma(1')+g_{\bar\sigma}(2')-g_\sigma(1)-g_{\bar\sigma}(2)\,]
\label{4.3}
\end{eqnarray}
(notation: $\bar\sigma=2$ for $\sigma=1$ and vice versa). The abbreviation $\delta_2(\ldots)$ means the delta-function $\delta(\epsilon_1+\epsilon_2-\epsilon_{1'}-\epsilon_{2'})$ with $\epsilon_1=k_1^2/2m$, etc. The pair-collision kernel $W^{(2)}_\sigma$ is written as
\begin{equation}
W^{(2)}_\sigma=2\pi|A_2|^2\,{1\over 4}\,\zeta_\sigma(1)\zeta_{\bar\sigma}(2)\zeta_\sigma(1')\zeta_{\bar\sigma}(2')~,
\label{4.4}
\end{equation}
where the scattering amplitude (at first order in interaction) $A_2=L^{-1}V_{12}(k_{1'}-k_1)\delta_{k_1+k_2,k_{1'}+k_{2'}}$, the Kronecker symbol $\delta_{k_1+k_2,k_{1'}+k_{2'}}$ signifies the total momentum conservation, $L$ is the system size, and $V_{12}(q)$ is the Fourier component of the interwire potential at the momentum $q$. The thermal factors $\zeta_\sigma$ are
\begin{equation}
\zeta_\sigma (1)=1/\!\cosh\, [\,(\epsilon_1-\epsilon_{F\sigma})/2T\,]~,\,\,{\rm etc.}
\label{4.5}
\end{equation}
For two-particle scattering, for which the energy and momentum conservation dictates that $k_2=k_{1'}$ and $k_{2'}=k_1$ (particles exchange momenta), the collision kernel $W^{(2)}_\sigma$ on the mass shell does not depend on $\sigma$ and
the collision integrals in two wires are related to each other by
\begin{equation}
{\rm st}^{(2)}_1(1)\zeta_1^2(1)=-{\rm st}^{(2)}_2(1)\zeta_2^2(1)
\label{4.6}
\end{equation}
for an arbitrary driving term and an arbitrary form of the collision kernel. Written explicitly, Eq.~(\ref{4.3}) reads
\begin{eqnarray}
{\rm st}^{(2)}_\sigma(k)\!\!&=&\!\! {m\over 2}\,\frac{\zeta_{\bar\sigma}(k)}{\zeta_\sigma(k)}
\int\!{dk'\over 2\pi}\,\zeta_\sigma(k')\zeta_{\bar\sigma}(k')\,
\frac{V_{12}^2(k'-k)}{|k'-k|}\nonumber\\
&\times&\!\!
\left[\,g_\sigma(k')+g_{\bar\sigma}(k)-g_\sigma(k)-g_{\bar\sigma}(k')\,\right]~.
\label{4.10a}
\end{eqnarray}

As a prelude to the solution of the kinetic equation, let us demonstrate the significance of the local relation (\ref{4.6}), specific to two-particle scattering, in rather general terms. The current in real space in wire $\sigma$ is given by
\begin{equation}
j_\sigma={e\over 8\pi m}\int\! dk\,k\zeta^2_\sigma g_\sigma~.
\label{4.18}
\end{equation}
Consider a steady-state situation in the dc limit with $j_1$ being finite (neither infinite nor zero) and $j_2=0$. For given $j_1$ and $j_2$, the functions $g_\sigma(k)$, which produce these currents according to Eq.~(\ref{4.18}), satisfy Eq.~(\ref{4.2}) with the vanishing term $-i\omega g_\sigma$ (for $j_\sigma$ held fixed with varying $\omega$, $g_\sigma$ is a regular function of $\omega$ in the dc limit, as explicitly demonstrated below). In view of Eq.~(\ref{4.6}), the fields $E_1$ and $E_2$ must, then, obey $E_1\zeta^2_1(k)=-E_2\zeta^2_2(k)$. For $\Delta\neq 0$, this relation can only be satisfied by
\begin{equation}
E_1=E_2=0, \qquad\quad \omega=0~,\quad\Delta\neq 0~.
\label{4.18a}
\end{equation}
As a consequence of that, the condition $j_2=0$ for $j_1\neq 0$ is maintained without applying any external electric field to wire 2 [to either wire for that matter, Eq.~(\ref{4.18a})]. This result (proven below by explicitly taking the limit $\omega\to 0$) is quite remarkable and means that the dc drag resistivity $\rho_{\rm D}=-(E_2/j_1)|_{j_2=0}$ vanishes for the case of two-particle scattering if $\Delta\neq 0$, as we already mentioned in Sec.~\ref{intro}. That is, two-particle scattering can only produce a finite $\rho_{\rm D}$ if the electron densities in the wires are the same.\cite{dmitriev12} At this point, it is worth emphasizing that it is crucial for the vanishing of $\rho_{\rm D}$ that the wires are assumed to be infinitely long and homogeneous, namely the (external) wavevector of the perturbation is sent to zero before taking the dc limit. This order of taking the limits defines the quantities known as the conductivity and the resistivity.

In turn, Eq.~(\ref{4.18a}) implies, when used in Eq.~(\ref{4.2}), that ${\rm st}_\sigma^{(2)}=0$. A unique property of two-particle scattering is that the collision integral (\ref{4.10a}) is nullified\cite{remark6} for $g_1(k)=g_2(k)$ [as shown below, this {\it is} the (only) solution to the kinetic equation with two-particle scattering for given $j_\sigma$ in the dc limit]. That is, for $\Delta\neq 0$, two-particle scattering necessarily leads to exact equilibration between the distribution functions $g_1$ and $g_2$ at the same momentum (here and below, we use the term ``distribution function" loosely to denote both $g_\sigma$ and $f_\sigma$; note that equal $g_\sigma$ for $\Delta\neq 0$ means different $f_\sigma$). Conversely, it is this sort of equilibration that is responsible for the cancelation between the incoming and outgoing terms in the collision integral. We conclude from this general discussion that there is an inherent link between the frictionless motion described by Eq.~(\ref{4.18a}) and the exact equilibration between $g_1$ and $g_2$, both peculiar to two-particle scattering.

Let us now turn to the solution of the kinetic equation (\ref{4.2}). As follows from Eq.~(\ref{4.6}), the kinetic equation has a zero-mode solution
\begin{eqnarray}
g_+\!\!&=&\!\!{1\over 2}\left(g_1{\zeta_1\over \zeta_2}+ g_2{\zeta_2\over\zeta_1}\right)
\label{4.7a}\\
&=&\!\!{ek\over 2mT}\left(E_1{\zeta_1\over \zeta_2}+E_2{\zeta_2\over\zeta_1}\right){1\over -i\omega}~,
\label{4.7b}
\end{eqnarray}
which is not subject to relaxation (here and below $-i\omega$ in the denominator is understood as $-i\omega+0$). The combination of $g_\sigma$ that relaxes to zero in the absence of the driving force is given by
\begin{equation}
g_-={1\over 2}(g_1-g_2)~.
\label{4.8}
\end{equation}
The closed equation for $g_-$ reads:
\begin{equation}
-i\omega g_--{ek\over 2mT}(E_1-E_2)={1\over 2}\left({\zeta_1\over\zeta_2}+{\zeta_2\over\zeta_1}\right){\rm st}_-~,
\label{4.9}
\end{equation}
where
\begin{eqnarray}
{\rm st}_-\!\!&=&\!\!{m\over 2}\int\!{dk'\over 2\pi}\,\zeta_1(k')\zeta_2(k')\,V_{12}^2(k'-k)\nonumber\\
&\times&\!\!{g_-(k')-g_-(k)\over |k'-k|}~.
\label{4.10}
\end{eqnarray}

For concreteness, let us focus on the Fokker-Planck limit, in which the characteristic momentum transfer $|k'-k|\sim 1/a$ in Eq.~(\ref{4.10}) is much smaller than $T/v_{F\sigma}$, with $v_{F\sigma}$ being the Fermi velocity in wire $\sigma$ and $a$ the characteristic spatial scale of the interwire potential. In this limit, the collision integral (\ref{4.10}) is written as
\begin{equation}
{\rm st}_-= 4{\cal D}^{(2)}\,{1\over \zeta_1\zeta_2}\,\partial_k\left(\zeta_1^2\zeta_2^2\,\partial_kg_-\right)~,
\label{4.11}
\end{equation}
where
\begin{equation}
{\cal D}^{(2)}={m\over 16}\int\!{dq\over 2\pi}\,|q|\,V^2_{12}(q)~.
\label{4.12}
\end{equation}
For identical wires, the constant ${\cal D}^{(2)}$ has the meaning of the diffusion coefficient in momentum space at the Fermi level.\cite{dmitriev12}
Integrating Eq.~(\ref{4.9}) at $\omega=0$ with ${\rm st}_-$ from Eq.~(\ref{4.11}), we have
\begin{eqnarray}
g_-&=&\!\!{e(E_1-E_2)\over 4mT {\cal D}^{(2)}}\int_0^k\!dk'\,{1\over\zeta^2_1(k')\zeta^2_2(k')}\nonumber\\
&\times&\!\!\int_{k'}^\infty\!dk''\,k''\,{\zeta_1^2(k'')\zeta_2^2(k'')\over \zeta_1^2(k'')+\zeta_2^2(k'')}~.
\label{4.13}
\end{eqnarray}
For $T\ll \epsilon_{F\sigma}$, two approximations in Eq.~(\ref{4.13}) are asymptotically accurate. First, the integral over $k''$ can be taken from 0 to $\infty$: it is then determined by $|k''|\simeq [m(\epsilon_{F1}+\epsilon_{F2})]^{1/2}$ and reduces to
\begin{equation}
\int_0^\infty \!\!dk\,k{\zeta^2_1\zeta^2_2\over \zeta^2_1+\zeta^2_2}\simeq 2mT\,{\cal I}\!\left({\Delta\over 2T}\right)~,\qquad T\ll\epsilon_{F\sigma}~,
\label{4.15a}
\end{equation}
where
\begin{equation}
{\cal I}(x)={\arctan (\tanh x)\over \sinh x}~.
\label{4.15}
\end{equation}
Second, the functions $\zeta_{1,2}(k')$ in the integral over $k'$ can be approximated as exponentials around $k'=0$: the integral over $k'$ is determined by $|k'|\sim \min\{|k|,(mT)^{1/2}\}$. For $g_-$ we thus obtain
\begin{eqnarray}
g_-\!\!&\simeq&\!\! {e(E_1-E_2)\over 64{\cal D}^{(2)}}\,(\pi mT)^{1/2}\,\Phi\!\left({k\over \sqrt{mT}}\right)\nonumber\\
&\times&\!\!{\cal I}\!\left({\Delta\over 2T}\right)\exp\!\left({\epsilon_{F1}+\epsilon_{F2}\over T}\right)~,
\label{4.14}
\end{eqnarray}
where $\Phi(x)=(2/\sqrt{\pi})\int_0^x\!dt\exp (-t^2)$ is the error function.

Writing $g_\sigma$ in terms of $g_\pm$, we separate $g_\sigma$ into a singular (at $\omega\to 0$) part, proportional to $g_+$, and a regular part, proportional to $g_-$:
\begin{eqnarray}
g_1\!\!&=&\!\!{2\zeta_1\zeta_2\over \zeta^2_1+\zeta^2_2}\left(g_++{\zeta_2\over\zeta_1}g_-\right)~,
\label{4.16}\\
g_2\!\!&=&\!\!{2\zeta_1\zeta_2\over \zeta^2_1+\zeta^2_2}\left(g_+-{\zeta_1\over\zeta_2}g_-\right)~.
\label{4.17}
\end{eqnarray}
Using $g_\pm$ from Eqs.~(\ref{4.7b}) and (\ref{4.14}) in Eqs.~(\ref{4.16}) and (\ref{4.17}), and substituting the resulting $g_\sigma$ in Eq.~(\ref{4.18}),
the conductivity matrix is written as
\begin{equation}
\hat\sigma={e^2\over m}{1\over -i\omega}
\left(\begin{array}{cc}
n_1 & 0 \\
0 & n_2 \\
\end{array}\right)-\sigma_{12}\left(\begin{array}{rr}
\!1 & \,-1 \\
\!-1 & \,1 \\
\end{array}\right)~,
\label{4.19}
\end{equation}
where
\begin{equation}
n_\sigma={1\over 8\pi mT}\int\!dk\,k^2\zeta^2_\sigma
\label{4.20}
\end{equation}
is the electron density in wire $\sigma$. For the off-diagonal term $\sigma_{12}$ we have (for $T\ll \epsilon_{F\sigma}$)
\begin{equation}
\sigma_{12}\simeq {e^2\bar{n}\over 2m}\,{\cal I}\!\left({\Delta\over 2T}\right)\left({1\over -i\omega}-{1\over \gamma}\right)+{\cal O}(\omega)~,
\label{4.21}
\end{equation}
where $\bar{n}=[(n_1^2+n_2^2)/2]^{1/2}$ and
\begin{equation}
\gamma\simeq {32\pi^{1/2}\bar{n}{\cal D}^{(2)}\over (mT)^{3/2}}\,{1\over {\cal I}\! \left(\Delta/ 2T\right)}\,\exp \left(-{\epsilon_{F1}+\epsilon_{F2}\over T}\right)~.
\label{4.22}
\end{equation}
The singular part of $\sigma_{12}$ is exactly given by $-e^2n/2im\omega$ for $n_1=n_2=n$. In the case of identical wires, the singular part of $\sigma_{12}$ is thus equal to that of $\sigma_{11}$. It is this property of the conductivity matrix $\hat\sigma$ that produces the nonzero resistivity matrix $\hat\rho=\hat\sigma^{-1}$ in the limit $\omega\to 0$ for two-particle scattering, calculated in Ref.~\onlinecite{dmitriev12}. Indeed, if $n_1\neq n_2$, the determinant of $\hat\sigma$ is seen to diverge as $1/\omega^2$ at $\omega\to 0$:
\begin{eqnarray}
\det\hat\sigma\simeq \!\!&-&\!\!\left({e^2\over m}\right)^2\left[\,n_1n_2-{n_1+n_2\over 2}\bar{n}\,{\cal I}\!\left({\Delta\over 2T}\right)\,\right]\,{1\over \omega^2}\nonumber\\
&+&\!\!{\cal O}\left(1/\omega\right)~.
\label{4.23}
\end{eqnarray}
As a result, $\hat\rho=0$ for $n_1\neq n_2$ and $\omega=0$: the finite damping (\ref{4.22}), which leads to the relaxation of $g_-$, is not sufficient to establish nonzero drag, mediated by two-particle scattering, between nonequivalent wires in the dc limit.\cite{remark7}

\begin{figure}
\centerline{\includegraphics[width=\columnwidth]{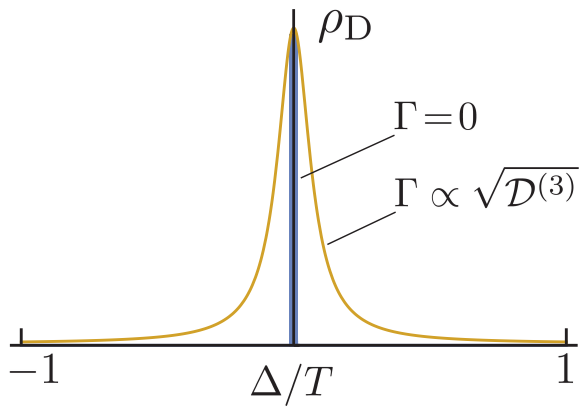}}
\caption{Schematic dependence of the drag resistivity $\rho_{\rm D}$ for two quantum wires on the chemical potential mismatch $\Delta$ (in units of the temperature $T$) for the case of weak intrawire equilibration. For two-particle scattering, $\rho_{\rm D}(\Delta)$ shows a bodyless resonance peak [Eq.~(\ref{4.24})] with a finite height and the width $\Gamma=0$. Three-particle scattering broadens the resonance [Eq.~(\ref{3.167})]: the width $\Gamma$ of the resulting Lorentzian scales with the strength of three-particle scattering ${\cal D}^{(3)}$ as $\sqrt{{\cal D}^{(3)}}$.}
\label{f1}
\end{figure}

For two-particle scattering, the dc drag resistivity $\rho_{\rm D}=-\rho_{12}(\omega=0)$ as a function of the mismatch $\Delta$ is thus a bodyless peak (Fig.~\ref{f1})---of a finite height (as obtained in Ref.~\onlinecite{dmitriev12}) and zero width:
\begin{eqnarray}
\rho_{\rm D}=\left\{
\begin{array}{ll}
\dfrac{m\gamma}{2e^2 n}~,\quad & \Delta=0~, \\ \\
0~, & \Delta\neq 0
\end{array}
\right.
\label{4.24}
\end{eqnarray}
[for $\Delta=0$, $\gamma$ from Eq.~(\ref{4.22}) coincides with $\gamma$ from Ref.~\onlinecite{dmitriev12}]. The derivation of Eq.~(\ref{4.24}) shows that the vanishing of $\rho_{\rm D}$ for $\Delta\neq 0$ is a generic property of two-particle scattering, independent of the particular form of the collision integral [the nonzero, for $\Delta\neq 0$, coefficient in front of $1/\omega^2$ in Eq.~(\ref{4.23}) comes from the contribution of $g_+$ to $\hat\sigma$]. This should be contrasted with the result of Ref.~\onlinecite{fuchs05}, where a finite resistivity for $\Delta\neq 0$ was obtained for the case of two-particle backscattering.\cite{remark5} This should also be contrasted with the results of Refs.~\onlinecite{pustilnik03} and \onlinecite{aristov07}, where Coulomb drag due to two-particle forward scattering was considered for both zero and nonzero $\Delta$, although forward scattering by itself does not produce nonzero $\rho_{\rm D}$, independently of whether $\Delta=0$ or not (see Ref.~\onlinecite{dmitriev12} for more detail).

Note that the significance of the function ${\cal I}(\Delta/2T)$ in the above is twofold. In Eq.~(\ref{4.22}), it reflects the dependence of the relaxation rate on $\Delta$ [which is of no importance to the behavior of $\rho_{\rm D}$ as a function of $\Delta$ in Eq.~(\ref{4.24})]. In Eq.~(\ref{4.23}), on the other hand, it comes from the modification of the singular (not related to relaxation at all) part of $\sigma_{12}$. The fact that the same function describes these two distinctly different aspects of the problem is actually specific to the use of the Fokker-Planck model: in general, the modification of the $T$ dependence of the regular part of $\sigma_{12}$ by finite $\Delta$ need not be described by Eq.~(\ref{4.21}) with $\gamma$ from Eq.~(\ref{4.22}). By contrast, the function ${\cal I}(\Delta/2T)$ in Eq.~(\ref{4.23}), and throughout Sec.~\ref{s2} below, is universal in the sense that it is independent of the particular form of the two-particle collision kernel.\cite{remark7}

It is also worth noting that the strong dependence of the damping (drag) rate on $\omega$ in the case of two-particle scattering, which was demonstrated for $\Delta=0$ in Ref.~\onlinecite{dmitriev12}, takes the extreme form for $\Delta\neq 0$. Expanding the solution of Eq.~(\ref{4.9}) to second order in $1/\omega$, the drag resistivity in the limit of large $\omega$ is obtained as
\begin{eqnarray}
&&\rho_{12}(\omega\to\infty)=-{m\over 32e^2n_1n_2T}\int\!{dk\over 2\pi}\,\zeta_1(k)\zeta_2(k)\nonumber\\
&&\times\!\int\!{dk'\over 2\pi}\,\zeta_1(k')\zeta_2(k')V^2_{12}(k'-k)|k'-k|~,
\label{4.25}
\end{eqnarray}
i.e., $\rho_{12}(\omega\to\infty)$ does not show any singularity at $\Delta=0$, whereas $\rho_{12}(\omega=0)$ vanishes if $\Delta\neq 0$ [Eq.~(\ref{4.24})]. If the damping rate was independent of $\omega$ (Drude-like ansatz), the drag resistivity would show no frequency dispersion.

To further elucidate the physics behind Eq.~(\ref{4.24}), let us assume that wires 1 and 2 are the active and passive wire, respectively, and write $j_1$ and $E_1$ under the condition that $j_2=0$ in terms of $E_2$ (the field in the passive wire necessary to maintain a currentless state therein):
\begin{eqnarray}
&&\hspace{-1cm}j_1={e^2\bar{n}\over m}\left[{A_b\over -i\omega} +{A_r\over \gamma}+{\cal O}(\omega)\right]E_2~,
\label{4.26}\\
&&\hspace{-1cm}E_1=\left[1-{2n_2\over \bar{n}{\cal I}(\Delta/2T)}\,\left(1-{i\omega\over\gamma}\right) +{\cal O}(\omega^2)\right]E_2~,
\label{4.27}
\end{eqnarray}
where the dimensionless constants $A_b$ and $A_r$ depend on $n_1/n_2$ and $\Delta/T$:
\begin{eqnarray}
&&A_b=-{2n_1n_2-(n_1+n_2)\bar{n}{\cal I}(\Delta/2T)\over \bar{n}^2{\cal I}(\Delta/2T)}~,\label{4.28}\\
&&A_r=-{2n_1n_2\over \bar{n}^2{\cal I}(\Delta/2T)}~.\label{4.29}
\end{eqnarray}
As seen from Eqs.~(\ref{4.26}) and (\ref{4.27}), if $j_1$ is held fixed with varying $\omega$, both $E_1$ and $E_2$ vanish as $\omega$ is decreased for $A_b\neq 0$, which is the case for $\Delta\neq 0$. This is because the ``relaxational" component of $j_1$ (proportional to $A_r$) is shunted by the ballistic component (proportional to $A_b$) in the dc limit. Specifically, to order ${\cal O}(\omega)$, the fields $E_1$ and $E_2$ for given $j_1$ and $j_2=0$ obey
\begin{eqnarray}
&&\hspace{-5mm}E_1={m\over e^2\bar{n}}\,{-i\omega\over A_b}\,\left[1-{2n_2\over\bar{n}{\cal I}(\Delta/2T)}\right]j_1 +{\cal O}(\omega^2)~,\label{4.30}\\
&&\hspace{-5mm}E_2={m\over e^2\bar{n}}\,{-i\omega\over A_b}\,j_1 +{\cal O}(\omega^2)
\label{4.31}
\end{eqnarray}
and do not depend on $\gamma$.

Substituting Eqs.~(\ref{4.30}) and (\ref{4.31}) in $g_+$ [Eq.~(\ref{4.7b})], the distribution function $g_\sigma$ [Eqs.~(\ref{4.16}) and (\ref{4.17})] for $j_2=0$ in the dc limit is given by
\begin{equation}
g_\sigma={1\over e\bar{n}T\!A_b}\,kF(k)\,j_1~,\qquad\qquad\omega=0~,
\label{4.32}
\end{equation}
where
\begin{equation}
F(k)={\zeta_1^2[\,1-2n_2/\bar{n}{\cal I}(\Delta/2T)\,]+\zeta^2_2\over \zeta^2_1+\zeta^2_2}~.
\label{4.33}
\end{equation}
Most importantly, as already mentioned below Eq.~(\ref{4.18a}), the distribution functions $g_1(k)$ and $g_2(k)$ at the same momentum $k$ equilibrate between themselves and become identical at $\omega\to 0$. Accordingly, the collision integral ${\rm st}^{(2)}_\sigma$ vanishes in the dc limit: the nonequilibrium steady state at $\omega\to 0$ is such that the incoming term in the collision integral is exactly canceled by the outgoing term. Crucially, the exact equilibration between $g_1$ and $g_2$ assumes that the dc limit is taken before $\Delta$ is sent to zero, which is why the case of $\Delta=0$ in Eq.~(\ref{4.24}) is special. Substituting Eq.~(\ref{4.32}) in Eq.~(\ref{4.18}) gives $j_2=0$ for a nonzero $j_1$. For $\Delta\neq 0$, Eq.~(\ref{4.32}) thus shows exactly how the dc current $j_2=0$ is maintained in wire 2 when one sends the dc current $j_1$ through wire 1. It is worth emphasizing once more that the distribution function (\ref{4.32}) is universal, i.e., does not depend on the particular form of the two-particle collision integral.

\section{Three-particle collision integral}
\label{s3}

As we will show below, in Sec.~\ref{s8a}, the singularity in the dependence of $\rho_{\rm D}$ on $\Delta$ in Eq.~(\ref{4.24}) is a peculiar property of two-particle scattering. Specifically, we will demonstrate that three-particle scattering leads to a broadening of the zero-width resonance of $\rho_{\rm D}$ at $\Delta=0$ and, if three-particle scattering becomes sufficiently strong, to a more conventional type of the dependence of $\rho_{\rm D}$ on $\Delta$ with a characteristic scale of $|\Delta|\sim T$. Let us therefore turn to the kinetic equation in the presence of three-particle scattering.

For arbitrary $\Delta$, the three-particle collision integral ${\rm st}^{(3)}_\sigma$, which should be added to the right-hand side of Eq.~(\ref{4.2}), is given by the sum
\begin{equation}
{\rm st}^{(3)}_\sigma={\rm st}^{(3a)}_\sigma+{\rm st}^{(3b)}_\sigma+{\rm st}^{(3c)}_\sigma~,
\label{3.2}
\end{equation}
where ${\rm st}^{(3a),(3b),(3c)}_\sigma$ are the contributions of scattering channels (a), (b), (c), respectively, defined as follows. For electron 1 in wire $\sigma$, which interacts with electrons 2 and 3, the scattering channels are defined (similar to Ref.~\onlinecite{dmitriev12}) depending on which wires electrons 2 and 3 belong to. Specifically, channel (a): all three electrons are in wire $\sigma$; channel (b): electrons 2 and 3 are in wire $\bar\sigma$; channel (c): electrons 2 and 3 are in different wires. The partial collision integrals at the momentum $k_1$ read
\begin{widetext}
\begin{eqnarray}
&&\hspace{-1cm}{\rm st}^{(3a)}_\sigma(1)={\eta_a\over \zeta^2_\sigma(1)}\!\sum_{231'2'3'}\!W^{(3a)}_\sigma\delta_3(\ldots)\,[\,g_\sigma(1')+g_\sigma(2')+g_\sigma(3')-g_\sigma(1)-g_\sigma(2)-g_\sigma(3)\,]~,
\label{3.3a}\\
&&\hspace{-1cm}{\rm st}^{(3b)}_\sigma(1)={\eta_b\over \zeta^2_\sigma(1)}\!\sum_{231'2'3'}\!W^{(3b)}_\sigma\delta_3(\ldots)\,
[\,g_\sigma(1')+g_{\bar\sigma}(2')+g_{\bar\sigma}(3')-g_\sigma(1)-g_{\bar\sigma}(2)-g_{\bar\sigma}(3)\,]~,
\label{3.3b}\\
&&\hspace{-1cm}{\rm st}^{(3c)}_\sigma(1)={\eta_c\over \zeta^2_\sigma(1)}\!\sum_{231'2'3'}\!W^{(3c)}_\sigma\delta_3(\ldots)\,[\,g_\sigma(1')+g_\sigma(2')+g_{\bar\sigma}(3')-g_\sigma(1)-g_\sigma(2)-g_{\bar\sigma}(3)\,]
\label{3.3c}
\end{eqnarray}
with $\eta_{a,b,c}=1/12\,,1/4\,,1/2$. Similar to Eq.~(\ref{4.3}), the coefficients in front of the functions $g_\sigma$ in the collision integrals are $\pm 1$, independently of whether $\Delta=0$ or not. The abbreviation $\delta_3(\ldots)$ stands for $\delta(\epsilon_1+\epsilon_2+\epsilon_3-\epsilon_{1'}-\epsilon_{2'}-\epsilon_{3'})$. The triple-collision kernels $W^{(3a),(3b),(3c)}_\sigma\propto \delta_{k_1+k_2+k_3,k_{1'}+k_{2'}+k_{3'}}$ are written as
\begin{eqnarray}
&&W^{(3a)}_\sigma(1',2',3'|1,2,3)=2\pi|A^{\rm irr}_{3a,\sigma}|^2\,{1\over 16}\,\zeta_\sigma(1)\zeta_\sigma(2)\zeta_\sigma(3)\zeta_\sigma(1')\zeta_\sigma(2')\zeta_\sigma(3')~,\label{3.5a}\\
&&W^{(3b)}_\sigma(1',2',3'|1,2,3)=2\pi|A^{\rm irr}_{3b,\sigma}|^2\,{1\over 16}\,\zeta_\sigma(1)\zeta_{\bar\sigma}(2)\zeta_{\bar\sigma}(3)\zeta_\sigma(1')\zeta_{\bar\sigma}(2')\zeta_{\bar\sigma}(3')~,\label{3.5b}\\
&&W^{(3c)}_\sigma(1',2',3'|1,2,3)=2\pi|A^{\rm irr}_{3c,\sigma}|^2\,{1\over 16}\,\zeta_\sigma(1)\zeta_\sigma(2)\zeta_{\bar\sigma}(3)\zeta_\sigma(1')\zeta_\sigma(2')\zeta_{\bar\sigma}(3')~,\label{3.5c}
\end{eqnarray}
\end{widetext}
with the irreducible (with respect to interaction) three-particle scattering amplitudes $A^{\rm irr}_{3a,\sigma}$, $A^{\rm irr}_{3b,\sigma}$, $A^{\rm irr}_{3c,\sigma}$ which are a direct generalization of those from Ref.~\onlinecite{dmitriev12} to the case of nonequal intrawire interaction potentials $V_{11}(q)$ and $V_{22}(q)$. If $V_{11}(q)=V_{22}(q)$, the subscript $\sigma$ in the amplitudes may be dropped. The amplitudes in Eqs.~(\ref{3.5a})-(\ref{3.5c}) are taken on the Slater determinants normalized to unity. Note that, generically, no local (for a given momentum) relation similar to Eq.~(\ref{4.6}) for two-particle scattering exists for three-particle scattering.

The coefficients $\eta_{a,b,c}$ account for the double counting of indistinguishable initial and final states in the unrestricted momentum summations in Eqs.~(\ref{3.3a})-(\ref{3.3c}), for more detail see Ref.~\onlinecite{dmitriev12} (note that the factors $\eta_{a,b,c}$ are missed in the formalism of Ref.~\onlinecite{lunde07}, which was used in a number of consequent publications). Each of the nonintegrable singularities\cite{dmitriev12} in the modulus squared of the three-particle scattering amplitudes (of the type $1/\Omega^2$ at zero energy $\Omega$ transferred in a virtual transition) must be regularized in the momentum summations in the collision integrals (\ref{3.3a})-(\ref{3.3c}) as the real part of a double pole or, equivalently, counterterms should be added to Eq.~(\ref{3.2}) to avoid a double counting of correlated two-particle collisions, see Sec.~IIIB in Ref.~\onlinecite{dmitriev12} and Refs.~\onlinecite{resibois65,bezzerides68} for a more detailed explanation of this point.

\section{Zero modes}
\label{s4}

The collision integrals (\ref{4.3}) for two-particle and (\ref{3.3a})-(\ref{3.3c}) for three-particle scattering, and those for an arbitrary number $N$ of colliding particles for that matter, are nullified for
\begin{equation}
g_\sigma=\Lambda_0+\Lambda_1k+\Lambda_2k^2~,\quad N\geq 2~,
\label{3.8}
\end{equation}
where $\Lambda_{0,1,2}$ are arbitrary $\sigma$-independent constants of $k$. The three terms in Eq.~(\ref{3.8}) correspond to the total number of particles, momentum, and energy conservation. The case of two-particle scattering is special: the collision integral (\ref{4.3}) is nullified for $g_\sigma$ given by an arbitrary $\sigma$-independent function of $k$, not only the polynomial (\ref{3.8}). Importantly, the total momentum conservation in the translation invariant system driven by the linear-in-$k$ inhomogeneous term in Eq.~(\ref{4.2}) does not necessarily imply that the singular at $\omega=0$ terms in $g_\sigma$, namely $g_\sigma^{\rm sing}\propto 1/\omega$, are also linear in $k$. This is quite generally the case for $\Delta=0$, where $g_\sigma^{\rm sing}=ek(E_1+E_2)/2mT(-i\omega)$ for arbitrary $N$. However, for $\Delta\neq 0$, the linearity of $g_\sigma^{\rm sing}$ in $k$ does not hold if only two-particle scattering is present: in that case, as follows from Eqs.~(\ref{4.16}) and (\ref{4.17}),
\begin{eqnarray}
g_\sigma^{\rm sing}\!\!&=&\!\!{2\zeta_1\zeta_2\over \zeta_1^2+\zeta_2^2}\,g_+\nonumber\\
&=&\!\!{ek\over mT}\,{E_1\zeta_1^2+E_2\zeta_2^2\over \zeta_1^2+\zeta_2^2}\,{1\over -i\omega}~,\quad N=2~,
\label{3.9}
\end{eqnarray}
and the linear combination of $g_1$ and $g_2$ that is time independent on all time scales is given by $g_+$ from Eq.~(\ref{4.7a}). If three-particle scattering is added, the degeneracy of the zero mode of the pair collision integral with respect to the dependence on $k$ is lifted, so that the odd-in-$k$ zero mode of the total (pair+triple) collision integral can only be given by Eq.~(\ref{3.8}) with $\Lambda_0=\Lambda_2=0$ and $\Lambda_1={\rm A}_1/(-i\omega)$:
\begin{equation}
g_\sigma^{\rm sing}={\rm A}_1{k\over -i\omega}~,\quad N\geq 3~,
\label{3.10}
\end{equation}
where ${\rm A}_1$ is a $k$ and $\sigma$ independent constant, i.e., can only be linear in $k$, and this is valid for any type of scattering beyond pair collisions. A nontrivial point to notice is that the residue of $g_\sigma^{\rm sing}$ at $\omega=0$ experiences a jump from the value given by Eq.~(\ref{3.9}) to the value given by Eq.~(\ref{3.10}) when the strength of three-particle scattering becomes nonzero [the discontinuity is only absent at  $\epsilon=(\epsilon_{F1}+\epsilon_{F2})/2$]. This means that the jump in $g_\sigma^{\rm sing}$ is of the form
\begin{eqnarray}
g_\sigma^{\rm sing}\!\!&=&\!\!\left({e\over mT}\,{E_1\zeta_1^2+E_2\zeta_2^2\over \zeta_1^2+\zeta_2^2}-{\rm A}_1\right)\nonumber\\
&\times&\!\!{k\over -i\omega+\Sigma_3(\omega,k)}
+{\rm A}_1{k\over -i\omega}~,
\label{3.11}
\end{eqnarray}
where $\Sigma_3(\omega,k)$ vanishes in the absence of three-particle scattering and has a finite value at $\omega=0$ in the presence of it.

To find ${\rm A}_1$ in Eq.~(\ref{3.10}), write the kinetic equation for $g_\sigma$ with ${\rm st}^{(3)}_\sigma\neq 0$ at $\omega\to 0$ as
\begin{equation}
\left({\rm A}_1-{eE_\sigma\over mT}\right)k={4\over \zeta_\sigma^2}\,\partial_kJ_\sigma~,
\label{3.12}
\end{equation}
where $J_\sigma(k)$ is the current in momentum space at the point $k$ in wire $\sigma$, related to the total time derivative of the momentum density $P_\sigma$ in wire $\sigma$ by
\begin{equation}
\dot{P}_\sigma=\int_{-\infty}^\infty\!{dk\over 2\pi}\,J_\sigma~.
\label{3.13}
\end{equation}
Multiplying Eq.~(\ref{3.12}) by $k\zeta^2_\sigma$ and integrating over all $k$, we have
\begin{eqnarray}
e(E_0-E_\sigma)n_\sigma=-\dot{P}_\sigma~,
\label{3.14a}
\end{eqnarray}
where the electron densities $n_\sigma$ are exactly given by Eq.~(\ref{4.20}) and
\begin{equation}
eE_0=mT{\rm A}_1~.
\label{3.15}
\end{equation}
The drag force $\dot{P}_\sigma$, acting on the unit length segment of wire $\sigma$, is thus compensated in the dc limit by the sum of the external force $-eE_\sigma n_\sigma$ and the zero-mode contribution to the force balance $eE_0n_\sigma$ that comes from the partial time derivative of $P_\sigma$. The effective electric field $E_0$ is obtained on quite general grounds solely as a result of the total momentum conservation $\dot{P}_1+\dot{P}_2=0$ (i.e., Newton's third law for the drag forces in two wires):
\begin{equation}
E_0={E_1n_1+E_2n_2\over n_1+n_2}~.
\label{3.16}
\end{equation}
Similar to Eq.~(\ref{3.11}), if only two-particle scattering is present, the force-balance equations at zero $\omega$ and $\Delta\neq 0$ are not of the form given by Eq.~(\ref{3.14a}); in particular, the effective zero-mode electric field coupled to the center-of-mass distribution ($E_0$ in the above) is different in two wires.

\section{Matrix structure}
\label{s5}

The singular (proportional to $1/\omega$) term  $j^{\rm sing}_\sigma=(e/8\pi m)\int\!dk\,k\zeta^2_\sigma g^{\rm sing}_\sigma$ in the current in wire $\sigma$ [with $g^{\rm sing}_\sigma$ from Eqs.~(\ref{3.10}), (\ref{3.15}), and (\ref{3.16})] gives the singular part $\hat\sigma^{\rm sing}$ of the conductivity matrix $\hat\sigma$:
\begin{equation}
\hat\sigma^{\rm sing}={e^2\over m}\,{1\over -i\omega}\,{1\over n_1+n_2}
\begin{pmatrix}
n_1^2 & n_1n_2 \\
n_1n_2 & n_2^2
\end{pmatrix}~.
\label{3.16a}
\end{equation}
The vanishing of the determinant
\begin{equation}
\det\hat\sigma^{\rm sing}=0
\label{3.16f}
\end{equation}
ensures that the elements of the resistivity matrix $\hat\rho=\hat\sigma^{-1}$ are generically nonzero in the dc limit. This is in contrast to the case of only two-particle scattering at $\Delta\neq 0$, where the determinant of the singular part of the conductivity matrix does not vanish [cf.\ Eq.~(\ref{4.23})]. Splitting the conductivity matrix into the singular and regular parts, $\hat\sigma=\hat\sigma^{\rm sing}+\hat\sigma^{\rm reg}$, the dc resistivity matrix for the case of $\det\hat\sigma^{\rm sing}=0$ is given by
\begin{equation}
\hat\rho(\omega=0)\equiv\lim_{\omega\to 0}\,{{\rm adj}\, \hat\sigma\over \det\hat\sigma}={{\rm adj}\,\hat\sigma^{\rm sing}\over {\rm Tr}\, [({\rm adj}\,\hat\sigma^{\rm sing})\,\hat\sigma^{\rm reg}]}~,
\label{3.16b}
\end{equation}
where the matrix structure of $\hat\rho(\omega=0)$ (regular at $\omega=0$) is identical to that of the adjugate of $\hat\sigma^{\rm sing}$ (singular at $\omega=0$),
\begin{equation}
{\rm adj}\,\hat\sigma^{\rm sing}={e^2\over m}\,{1\over -i\omega}\,{1\over n_1+n_2}
\begin{pmatrix}
n_2^2 & -n_1n_2 \\
-n_1n_2 & n_1^2
\end{pmatrix}~,
\label{3.16c}
\end{equation}
with
\begin{eqnarray}
&&\hspace{-3mm}{\rm Tr}\, [({\rm adj}\,\hat\sigma^{\rm sing})\,\hat\sigma^{\rm reg}]={e^2\over m}\,{1\over -i\omega}\,{1\over n_1+n_2}\nonumber\\
&&\hspace{-3mm}\times\left[\,n_2^2\sigma^{\rm reg}_{11}+n_1^2\sigma^{\rm reg}_{22}-n_1n_2\left(\sigma^{\rm reg}_{12}+\sigma^{\rm reg}_{21}\right)\,\right]~.
\label{3.16d}
\end{eqnarray}
The dc drag resistivity $\rho_{\rm D}=-\rho_{12}(\omega=0)$ is thus obtained as
\begin{equation}
\rho_{\rm D}={n_1n_2\over n_2^2\sigma^{\rm reg}_{11}+n_1^2\sigma^{\rm reg}_{22}-n_1n_2\left(\sigma^{\rm reg}_{12}+\sigma^{\rm reg}_{21}\right)}~.
\label{3.16e}
\end{equation}
Note that the $\hat\rho(\omega=0)$ matrix is singular (zero determinant),
\begin{equation}
\det\hat\rho(\omega=0)=0~,
\label{3.16g}
\end{equation}
and symmetric (with generically different diagonal entries, i.e., the structure of the matrix is characterized by two independent parameters, e.g., $n_1$ and $n_2$). The matrix structure of $\hat\sigma^{\rm reg}$ will be derived in Eq.~(\ref{3.152}) below.

\vspace{5mm}
\section{Fokker-Planck equation}
\label{s6}

The crucial difference between the solutions of the kinetic equation in the dc limit for $\Delta=0$ and $\Delta\neq 0$ is that in the latter case three-particle scattering cannot be neglected in the calculation of $\rho_{\rm D}$ (even in the limit of an infinitesimally small but finite strength of three-particle scattering), because otherwise $\rho_{\rm D}=0$ [Eq.~(\ref{4.24})]. We turn now to the solution of the kinetic equation in the presence of both two- and three-particle scattering. Similar to Sec.~\ref{s2}, we focus on the Fokker-Planck formulation of the kinetic problem.
The derivation of the Fokker-Planck collision integral in the presence of three-particle scattering, with a discussion of important peculiarities of three-particle scattering compared to two-particle scattering (in regard to the selection of momenta contributing to the coefficients of the gradient expansion), is presented in Appendix \ref{aB}. The contribution to the current in momentum space $J_\sigma$ in wire $\sigma$ of three-particle scattering, $J^{(3)}_\sigma$, in the Fokker-Planck limit is written as
\begin{equation}
J^{(3)}_\sigma(k)=D^{(3)}_\sigma(k)\partial g_\sigma(k)-C_\sigma(k)~,
\label{3.39z}
\end{equation}
where the diffusion coefficient in momentum space $D^{(3)}_\sigma=D_\sigma^{(3a)}+D_\sigma^{(3b)}+D_\sigma^{(3c)}$ and the integral term $C_\sigma=C^{(3a)}_\sigma+C^{(3b)}_\sigma+C^{(3c)}_\sigma$ are sums of terms coming from channels (a), (b), and (c):
\begin{widetext}
\vspace{-5mm}
\begin{eqnarray}
&&D_\sigma^{(3a),(3b),(3c)}(k_1)={1\over 8\nu_{a,b,c}}\sum_{231'2'3'}\,\!\!\!\!\!^d \,W_\sigma^{(3a),(3b),(3c)}(1',2',3'|1,2,3)\delta_3(\ldots)q^2~,
\label{3.83d}\\
&&C_\sigma^{(3a)}(k_1)=-{1\over 4\nu_a}\sum_{231'2'3'}\,\!\!\!\!\!^d \,W_\sigma^{(3a)}(1',2',3'|1,2,3)\delta_3(\ldots)qq_2\,\partial_{k_2}g_\sigma(2)~,
\label{3.83a}\\
&&C_\sigma^{(3b)}(k_1)=-{1\over 4\nu_b}\sum_{231'2'3'}\,\!\!\!\!\!^d \,W_\sigma^{(3b)}(1',2',3'|1,2,3)\delta_3(\ldots)qq_2\,\partial_{k_2}g_{\bar\sigma}(2)~,\label{3.83b}\\
&&C_\sigma^{(3c)}(k_1)=-{1\over 8\nu_c}\sum_{231'2'3'}\,\!\!\!\!\!^d \,W_\sigma^{(3c)}(1',2',3'|1,2,3)\delta_3(\ldots)q\left[\,q_2\,\partial_{k_2}g_\sigma(2)+q_3\,\partial_{k_3}g_{\bar\sigma}(3)\,\right]
\label{3.83c}
\end{eqnarray}
with $\nu_{a,b,c}=2,2,1$ (for more detail on the combinatorial factors $\nu_{a,b,c}$ see Appendix~\ref{aB}). The sign $^d$ in $\sum^d$ restricts the summation to direct scattering (as opposed to exchange processes) with $|k_1-k_{1'}|,|k_2-k_{2'}|,|k_3-k_{3'}|\alt 1/a$. Note that the cases of two- and three-particle scattering are different in that, in the latter case, there appears the integral term $C_\sigma$ in the Fokker-Planck expression for the current in momentum space.\cite{remark3}

The resulting Fokker-Planck equation in the dc limit reads:
\begin{equation}
{e\over 4mT}\,{n_{\bar\sigma}\over n_\sigma+n_{\bar\sigma}}(E_{\bar\sigma}-E_\sigma)k={1\over\zeta^2_\sigma}\,
\partial_k\!
\left\{\zeta^2_\sigma\!\left[\,{1\over 2}{\cal D}^{(2)}\zeta^2_{\bar\sigma}\,\partial_k({\rm g}_\sigma-{\rm g}_{\bar\sigma})
+{\cal D}^{(3)}_\sigma(k)\partial_k{\rm g}_\sigma-{\cal C}_\sigma(k)\right]\right\},
\label{3.40}
\end{equation}
\end{widetext}
where
\begin{equation}
{\rm g}_\sigma=g_\sigma-g^{\rm sing}_\sigma
\label{3.40a}
\end{equation}
is the regular part of $g_\sigma$ for $\omega=0$ and we introduced the functions
${\cal D}^{(3)}_\sigma(k)$ and ${\cal C}_\sigma(k)$ by explicitly extracting the factors $\zeta^2_\sigma(k)$, sharply peaked on the Fermi surface, from $D^{(3)}_\sigma(k)$ and $C_\sigma(k)$, with the partial terms in ${\cal D}^{(3)}_\sigma(k)$ and ${\cal C}_\sigma(k)$ being related to those in Eqs.~(\ref{3.83d})-(\ref{3.83c}) by
\begin{eqnarray}
{\cal D}_\sigma^{(3a),(3b),(3c)}(k)\!\!&=&\!\!D_\sigma^{(3a),(3b),(3c)}(k)/\zeta^2_\sigma(k)~,\label{3.40b}\\
{\cal C}_\sigma^{(3a),(3b),(3c)}(k)\!\!&=&\!\!C_\sigma^{(3a),(3b),(3c)}(k)/\zeta^2_\sigma(k)~.\label{3.40c}
\end{eqnarray}
Importantly, Eq.~(\ref{3.40}) for $\Delta\neq 0$ does not have a solution if ${\cal D}_\sigma^{(3)}$ and ${\cal C}_\sigma$ are zero. This is because the dc limit in Eq.~(\ref{3.40}) is assumed to be taken before the limit ${\cal D}_\sigma^{(3)}\to 0$ and ${\cal C}_\sigma\to 0$.

Multiplying Eq.~(\ref{3.40}) by $\zeta^2_\sigma$ and integrating over $k$, we have
\begin{eqnarray}
&&{e\over 2}\,{n_{\bar\sigma}\over n_\sigma+n_{\bar\sigma}}(E_\sigma-E_{\bar\sigma})(1-\tanh_\sigma)\nonumber\\
&&=\zeta^2_\sigma\left[\,{1\over 2}{\cal D}^{(2)}\zeta^2_{\bar\sigma}\,\partial_k({\rm g}_\sigma-{\rm g}_{\bar\sigma})+{\cal D}_\sigma^{(3)}(k)\partial_k{\rm g}_\sigma
-{\cal C}_\sigma(k)\,\right]~,\nonumber\\
\label{3.110}
\end{eqnarray}
where
\begin{equation}
\tanh_\sigma=\tanh {k^2-k_{F\sigma}^2\over 4mT}
\label{3.111}
\end{equation}
and $k_{F\sigma}$ is the Fermi momentum in wire $\sigma$. Solving the two-component ($\sigma=1,2$) integro-differential equation (\ref{3.110}) for $\partial_k{\rm g}_\sigma$ in terms of the integrals ${\cal C}_\sigma(k)$ gives
\begin{equation}
\partial_k{\rm g}_\sigma={{\cal D}^{(2)}\zeta^2_\sigma\zeta^2_{\bar\sigma}(h_\sigma+h_{\bar\sigma})/2+{\cal D}_{\bar\sigma}^{(3)}\zeta^2_{\bar\sigma}h_\sigma\over \zeta^2_\sigma\zeta^2_{\bar\sigma}\,{\textsf D}}~,
\label{3.114}
\end{equation}
where
\begin{equation}
h_\sigma={e\over 2}\,{n_{\bar\sigma}\over n_\sigma +n_{\bar\sigma}}(E_\sigma-E_{\bar\sigma})(1-\tanh_\sigma)+\zeta^2_\sigma {\cal C}_\sigma
\label{3.115}
\end{equation}
and
\begin{equation}
{\textsf D}={1\over 2}{\cal D}^{(2)}\!\left[\,{\cal D}_\sigma^{(3)}\zeta^2_\sigma+{\cal D}_{\bar\sigma}^{(3)}\zeta^2_{\bar\sigma}\,\right]+{\cal D}^{(3)}_\sigma
{\cal D}^{(3)}_{\bar\sigma}~.
\label{3.55}
\end{equation}
Note that, unless $\Delta=0$ (i.e., $h_\sigma=-h_{\bar\sigma}$), $\partial_k{\rm g}_\sigma$ in Eq.~(\ref{3.114}) diverges in the limit of vanishing three-particle scattering [see the comment at the end of the paragraph below Eq.~(\ref{3.40a})].

Summing up the two components of Eq.~(\ref{3.110}) [thus excluding the terms proportional to ${\cal D}^{(2)}$] and integrating the sum over all $k$, one obtains the total momentum-conservation law for triple collisions in the form
\begin{equation}
\sum_\sigma\int\!dk\,\zeta^2_\sigma\left[\,{\cal D}_\sigma^{(3)}(k)\partial_k{\rm g}_\sigma
-{\cal C}_\sigma(k)\,\right]=0~.
\label{3.112}
\end{equation}
For given ${\cal D}_\sigma^{(3)}(k)$, any model approximation for ${\cal C}_\sigma(k)$ must obey Eq.~(\ref{3.112}). Note that,
as shown in Appendix~\ref{aC}, there are also ``partial" conservation laws [see Eqs.~(\ref{3.70a}) and (\ref{3.70b})] which require, additionally, that
\begin{eqnarray}
&&\hspace{-10mm}\int\!dk\,\zeta^2_\sigma\left[\,{\cal D}_\sigma^{(3a)}(k)\partial_k{\rm g}_\sigma
-{\cal C}_\sigma^{(3a)}(k)\,\right]=0~,
\label{3.113a}\\
&&\hspace{-10mm}\int\!dk\,\zeta^2_\sigma\left[\,{\cal D}_\sigma^{(3b)}(k)\partial_k{\rm g}_\sigma
-{\cal C}_\sigma^{(3b)}(k)\,\right]\nonumber\\
&&\hspace{-10mm}+\int\!dk\,\zeta^2_{\bar\sigma}\left[\,{\cal D}_{\bar\sigma}^{(3c)}(k)\partial_k{\rm g}_{\bar\sigma}
-{\cal C}_{\bar\sigma}^{(3c)}(k)\,\right]=0~.
\label{3.113b}
\end{eqnarray}
While Eq.~(\ref{3.113a}) has exactly the same meaning for a single wire as Eq.~(\ref{3.112}) has for two, Eq.~(\ref{3.113b}) links channels (b) and (c) to each other through the ``integral" constraint.

\section{Exactly solvable Fokker-Planck equation}
\label{s7}

Turning to a very instructive example of the Fokker-Planck description of three-particle scattering, let us solve an exactly solvable model with ${\cal D}^{(3)}_\sigma$ and ${\cal C}_\sigma$ being independent of $k$. These should be chosen to satisfy Eq.~(\ref{3.112}), which gives
\begin{equation}
\big[\,{\cal D}^{(3)}_\sigma\langle\partial_k{\rm g}_\sigma\rangle-{\cal C}_\sigma\,\big]k_{T\sigma}=-\big[\,{\cal D}^{(3)}_{\bar\sigma}\langle\partial_k{\rm g}_{\bar\sigma}\rangle-{\cal C}_{\bar\sigma}\,\big]k_{T\bar\sigma}~,
\label{3.116a}
\end{equation}
where
\begin{equation}
\left\langle\partial_k{\rm g}_\sigma\right\rangle=\left\langle\partial_k{\rm g}_\sigma\right\rangle_\sigma~,
\label{3.116b}
\end{equation}
the average $\left\langle\ldots\right\rangle_\sigma$ is defined as
\begin{equation}
\left\langle\ldots\right\rangle_\sigma={1\over k_{T\sigma}}\int\!dk\,\zeta^2_\sigma\,(\ldots)~,
\label{3.116c}
\end{equation}
and
\begin{equation}
k_{T\sigma}=\int\!dk\,\zeta^2_\sigma=8\pi T(\partial n/\partial\mu)_\sigma\simeq 8mT/k_{F\sigma}~,
\label{3.116d}
\end{equation}
with $(\partial n/\partial\mu)_\sigma$ the thermodynamic density of states (compressibility) in wire $\sigma$. If one takes, further, ${\cal C}_\sigma$ to be a linear function of the averages (\ref{3.116b}),
\begin{equation}
{\cal C}_\sigma=\lambda_\sigma\langle\partial_k{\rm g}_\sigma\rangle + \mu_\sigma\langle\partial_k{\rm g}_{\bar\sigma}\rangle
\label{3.116}
\end{equation}
with arbitrary constants $\lambda_\sigma$ and $\mu_\sigma$, then Eq.~(\ref{3.116a}), which should be satisfied for arbitrary ${\rm g}_\sigma(k)$ [on the model level: on the space of ${\rm g}_\sigma(k)$ that supports the representation (\ref{3.116})], requires that ${\cal D}^{(3)}_\sigma$ and ${\cal C}_\sigma$ be related to each other by
\begin{equation}
{\cal D}^{(3)}_\sigma=\lambda_\sigma+\mu_{\bar\sigma}k_{T\bar\sigma}/k_{T\sigma}~.
\label{3.117}
\end{equation}
If three-particle scattering is only present in channel (a), this can be modeled by putting $\mu_\sigma=0$ [cf.\ Eq.~(\ref{3.113a})].

Multiplying Eq.~(\ref{3.114}) by $\zeta^2_\sigma$, substituting Eq.~(\ref{3.116}) for ${\cal C}_\sigma$ in $h_\sigma$, and integrating over all $k$, we have a $2\times 2$ algebraic equation for $\langle\partial_k{\rm g}_\sigma\rangle$:
\begin{equation}
a_\sigma\langle\partial_k{\rm g}_\sigma\rangle+b_\sigma\langle\partial_k{\rm g}_{\bar\sigma}\rangle =c_\sigma~,
\label{3.53}
\end{equation}
where
$a_\sigma=\langle\tilde{a}_\sigma\rangle_\sigma$, $b_\sigma=\langle\tilde{b}_\sigma\rangle_\sigma$, $c_\sigma=\langle\tilde{c}_\sigma\rangle_\sigma$, and
\begin{eqnarray}
&&\hspace{-5mm}\tilde{a}_\sigma=1-{1\over 2{\textsf D}}\left[\,{\cal D}^{(2)}\left(\lambda_\sigma\zeta^2_\sigma+\mu_{\bar\sigma}\zeta^2_{\bar\sigma}\right)+2{\cal D}^{(3)}_{\bar\sigma}\lambda_\sigma\right],
\label{3.54a}\\
&&\hspace{-5mm}\tilde{b}_\sigma=-{1\over 2{\textsf D}}\left[\,{\cal D}^{(2)}\left(\mu_\sigma\zeta^2_\sigma+\lambda_{\bar\sigma}\zeta^2_{\bar\sigma}\right)+2{\cal D}^{(3)}_{\bar\sigma}\mu_\sigma\right],
\label{3.54b}\\
&&\hspace{-5mm}\tilde{c}_\sigma={e(E_\sigma-E_{\bar\sigma})\over 4(n_\sigma+n_{\bar\sigma})} {1\over {\textsf D}}\bigg\{\,{\cal D}^{(2)}\!\left[\,n_{\bar\sigma}(1-\tanh_\sigma)\right.\nonumber\\
&&\left.\left.\hspace{-1.5mm}-n_\sigma(1-\tanh_{\bar\sigma})\,\right]+2{\cal D}^{(3)}_{\bar\sigma}n_{\bar\sigma}{1-\tanh_\sigma\over \zeta^2_\sigma}\,\right\}.
\label{3.54c}
\end{eqnarray}
Substituting the solution of Eq.~(\ref{3.53}) for $\langle\partial_k{\rm g}_\sigma\rangle$ in Eq.~(\ref{3.114}) solves the integral equation (\ref{3.114}) for $\partial_k{\rm g}_\sigma$, within the model specified in Eq.~(\ref{3.116}).

This brings us to a subtle but crucially important point about the constraints, related to momentum conservation, on ${\cal C}_\sigma$ within the model (\ref{3.116}). Specifically, the question is how many independent constants can be chosen to parametrize the model of three-particle scattering, formulated in terms of six constants ${\cal D}^{(3)}_\sigma$, $\lambda_\sigma$, and $\mu_\sigma$. The total momentum-conservation law (\ref{3.112}), applied to the model (\ref{3.116}) in the form of Eq.~(\ref{3.117}), leaves four of them independent,\cite{remark2} e.g., $\lambda_\sigma$ and $\mu_\sigma$. Importantly, however, these are not independent. In fact, the model (\ref{3.116}) has additional constraints and is parametrized by two independent constants for arbitrary $\Delta$ [for the solution to Eqs.~(\ref{3.114}) and (\ref{3.53}) in the limit $\Delta\to 0$, which reproduces the result of Ref.~\onlinecite{dmitriev12}, see Appendix~\ref{aD}].

The additional [compared to Eq.~(\ref{3.117}) or, more generally, Eq.~(\ref{3.112})] constraint is also a direct consequence of momentum conservation. Namely, the collision integral for arbitrary $\Delta$ must be nullified for
\begin{equation}
g_\sigma(k)=\Lambda_1 k~,
\label{3.135a}
\end{equation}
where the (arbitrary) constant $\Lambda_1$ is independent of $\sigma$ [cf.\ Eqs.~(\ref{3.8}) and (\ref{3.10}) in Sec.~\ref{s4}]. In the Fokker-Planck formulation, this means that ${\cal C}_\sigma(k)=\Lambda_1{\cal D}_\sigma^{(3)}(k)$ for $\partial_kg_\sigma=\partial_kg_{\bar\sigma}=\Lambda_1$. For the model from Eq.~(\ref{3.116}), this condition translates into
\begin{equation}
{\cal D}^{(3)}_\sigma=\lambda_\sigma+\mu_\sigma~.
\label{3.135}
\end{equation}
Combining Eqs.~(\ref{3.135}) and (\ref{3.117}), we obtain the relation between the coefficients in Eq.~(\ref{3.116}) for different $\sigma$ that must be satisfied as a result of the existence of the zero mode (\ref{3.135a}):
\begin{equation}
\mu_\sigma k_{T\sigma}=\mu_{\bar\sigma} k_{T\bar\sigma}
\label{3.136}
\end{equation}
(with no constraint on $\lambda_\sigma$). Substituting Eq.~(\ref{3.136}) in Eqs.~(\ref{3.54a}) and (\ref{3.54b}), we have
\begin{equation}
a_\sigma=-b_\sigma~.
\label{3.137}
\end{equation}
Equation (\ref{3.137}) can also be obtained by requiring that the left-hand side of Eq.~(\ref{3.53}) vanishes for the homogeneous solution (\ref{3.135a}).

The existence of the zero mode (\ref{3.135a}) associated with momentum conservation has the consequence that the determinant of Eq.~(\ref{3.53}) is zero,
\begin{equation}
a_\sigma a_{\bar\sigma}-b_\sigma b_{\bar\sigma}=0~,
\label{3.118a}
\end{equation}
for arbitrary $\Delta$, not only $\Delta=0$ (cf.\ Appendix~\ref{aD}). As a result of the degeneracy of Eq.~(\ref{3.53}), we have
\begin{equation}
\langle\partial_k{\rm g}_\sigma\rangle-\langle\partial_k{\rm g}_{\bar\sigma}\rangle=-c_\sigma/b_\sigma~,
\label{3.138}
\end{equation}
with
\begin{equation}
c_\sigma/b_\sigma=-c_{\bar\sigma}/b_{\bar\sigma}~.
\label{3.139}
\end{equation}
Equation (\ref{3.139}) thus gives one more, in addition to Eq.~(\ref{3.136}), relation between the coefficients in Eq.~(\ref{3.116}). Altogether, we have four algebraic constraints on six constants ${\cal D}^{(3)}_\sigma$, $\lambda_\sigma$, and $\mu_\sigma$ that parametrize three-particle scattering for given $\Delta$ [Eqs.~(\ref{3.117}), (\ref{3.136}), and (\ref{3.139})], i.e., one can choose only two constants arbitrarily [similar to the case $\Delta=0$, where the model is parametrized by ${\cal D}^{(3)}=\lambda+\mu$ and $\lambda-\mu$, see Appendix~\ref{aD}]. Note that Eq.~(\ref{3.136}) fixes the relation between $\mu_1$ and $\mu_2$ only, thus being related to the partial conservation law (\ref{3.113b}). The important point to notice is that Eq.~(\ref{3.112}), reflecting the total momentum-conservation law per se (without a reference to the partial conservation laws), and the vanishing of the collision integral for the associated zero mode (\ref{3.135a}) are not equivalent in the sense of the constraints they require for the parameters of the model (\ref{3.116}).

Using Eq.~(\ref{3.139}), one can represent Eq.~(\ref{3.114}) for the model (\ref{3.116}) in the form
\begin{equation}
\partial_k{\rm g}_\sigma=\langle\partial_k{\rm g}_\sigma\rangle+{1\over b_\sigma}\left(b_\sigma\tilde{c}_\sigma -\tilde{b}_\sigma c_\sigma\right)~.
\label{3.140}
\end{equation}
Together with Eq.~(\ref{3.138}), this solves Eq.~(\ref{3.110}) in the model (\ref{3.116}) for the difference ${\rm g}_\sigma-{\rm g}_{\bar\sigma}$:
\begin{equation}
\partial_k({\rm g}_\sigma-{\rm g}_{\bar\sigma})=-{c_\sigma\over b_\sigma}\left(1+\tilde{b}_\sigma +\tilde{b}_{\bar\sigma}\right)+\tilde{c}_\sigma-\tilde{c}_{\bar\sigma}~,
\label{3.142}
\end{equation}
with
\begin{equation}
1+\tilde{b}_\sigma +\tilde{b}_{\bar\sigma}={1\over {\textsf D}}\left(\lambda_\sigma\lambda_{\bar\sigma}-\mu_\sigma\mu_{\bar\sigma}\right)
\label{3.143}
\end{equation}
and
\begin{eqnarray}
&&\hspace{-10mm}\tilde{c}_\sigma-\tilde{c}_{\bar\sigma}={e(E_\sigma -E_{\bar\sigma})\over 2(n_\sigma+n_{\bar\sigma})\textsf{D}}\nonumber\\
&&\hspace{-10mm}\times\left[{\cal D}^{(3)}_\sigma n_\sigma{1-\tanh_{\bar\sigma}\over\zeta^2_{\bar\sigma}}+{\cal D}^{(3)}_{\bar\sigma}n_{\bar\sigma}{1-\tanh_\sigma\over\zeta^2_\sigma}\right]~.
\label{3.144}
\end{eqnarray}
For $\Delta=0$, Eq.~(\ref{3.142}) reduces to Eq.~(\ref{3.46a}) (with $\tilde{b}_\sigma=\tilde{b}_{\bar\sigma}$ and $\tilde{c}_\sigma=-\tilde{c}_{\bar\sigma}$).

The Fokker-Planck equation in the form of Eq.~(\ref{3.40}) [or Eq.~(\ref{3.114})] is fully solvable only for the difference $\partial_k({\rm g}_\sigma-{\rm g}_{\bar\sigma})$ [Eq.~(\ref{3.142})] and is degenerate with respect to a shift of $\partial_k{\rm g}_\sigma$ by a $k$ and $\sigma$ independent constant (``homogeneous solution"), unless a boundary condition lifts the degeneracy (which is not the case here). That is, the function ${\rm g}_\sigma+{\rm g}_{\bar\sigma}$, obeying [from Eqs.~(\ref{3.138}) and (\ref{3.140})]
\begin{eqnarray}
\partial_k({\rm g}_\sigma+{\rm g}_{\bar\sigma})\!\!&=&\!\!\langle\partial_k{\rm g}_\sigma\rangle+\langle\partial_k{\rm g}_{\bar\sigma}\rangle
+\tilde{c}_\sigma+\tilde{c}_{\bar\sigma}\nonumber\\
&+&\!\!(\tilde{b}_\sigma-\tilde{b}_{\bar\sigma})(\langle\partial_k{\rm g}_\sigma\rangle-\langle\partial_k{\rm g}_{\bar\sigma}\rangle)~,
\label{3.146}
\end{eqnarray}
is obtainable from Eq.~(\ref{3.40}) up to the constant
\begin{equation}
B=\langle\partial_k{\rm g}_\sigma\rangle+\langle\partial_k{\rm g}_{\bar\sigma}\rangle~.
\label{3.146a}
\end{equation}
In particular, this means that the inhomogeneous solution of Eq.~(\ref{3.40}), which depends on the fields $E_\sigma$ only through the combination $E_\sigma-E_{\bar\sigma}$, namely
\begin{equation}
{\rm g}_\sigma(k)\propto E_\sigma-E_{\bar\sigma}~,
\label{3.146b}
\end{equation}
can be shifted as
\begin{equation}
{\rm g}_\sigma(k)\to{\rm g}_\sigma(k)+\vartheta_\sigma (E_\sigma-E_{\bar\sigma})k
\label{3.145}
\end{equation}
with an arbitrary constant $\vartheta_1=-\vartheta_2$.

Note, however, that the $k$ and $\sigma$ independent term in $\partial_k{\rm g}_\sigma$ is of no importance if one is only interested in finding $\rho_{\rm D}$ in Eq.~(\ref{3.16e}). This is because $\rho_{\rm D}$ does not change after the transformation (\ref{3.145}). That is, while the conductivity matrix $\hat\sigma^{\rm reg}$ depends on $\vartheta_\sigma$, being shifted by the $\vartheta_\sigma$ dependent term as
\begin{equation}
\hat\sigma^{\rm reg}\to \hat\sigma^{\rm reg}+eT\vartheta_1\left(\begin{array}{rr}
n_1 & \,-n_1 \\
n_2 & \,-n_2 \\
\end{array}\right)~,
\label{3.145a}
\end{equation}
the resistivity $\rho_{\rm D}$ does not [in fact, $\rho_{\rm D}$ is not changed after a more general transformation of the form ${\rm g}_\sigma\to{\rm g}_\sigma+(\vartheta_\sigma E_\sigma +\vartheta_{\bar\sigma} E_{\bar\sigma})k$ with two arbitrary constants $\vartheta_\sigma$].

The constant $B$ [Eq.~(\ref{3.146a})] is found by retaining the finite-$\omega$ term $-i\omega {\rm g}_\sigma$ [with ${\rm g}_\sigma$ defined in Eq.~(\ref{3.40a})] on the left-hand side of Eq.~(\ref{3.12}) (which was written for $\omega\to 0$) and repeating the steps that led to Eq.~(\ref{3.16}). In addition to the condition (\ref{3.16}) for the singular part of $g_\sigma$, we have now one more condition for the regular part, namely
\begin{equation}
\int\!dk\left(\zeta^2_\sigma {\rm g}_\sigma + \zeta^2_{\bar\sigma}{\rm g}_{\bar\sigma}\right)k=0~,
\label{3.147}
\end{equation}
or, equivalently,
\begin{equation}
\int\!dk\left[\,\left(1-\tanh_\sigma\right)\partial_k{\rm g}_\sigma+\left(1-\tanh_{\bar\sigma}\right)\partial_k{\rm g}_{\bar\sigma}\,\right]=0~.
\label{3.147a}
\end{equation}
Equation (\ref{3.147}) means that the regular part $j^{\rm reg}_\sigma=(e/8\pi m)\int\!dk\,k\zeta^2_\sigma {\rm g}_\sigma$ of the current in real space in wire $\sigma$ obeys the relation
\begin{equation}
j^{\rm reg}_\sigma=-j^{\rm reg}_{\bar\sigma}~,
\label{3.147b}
\end{equation}
in contrast to the singular part $j^{\rm sing}_\sigma$, which, according to Eq.~(\ref{3.10}), obeys
\begin{equation}
j^{\rm sing}_\sigma n_{\bar\sigma}=j^{\rm sing}_{\bar\sigma} n_\sigma~.
\label{3.147c}
\end{equation}
We thus obtain, from Eqs.~(\ref{3.138}), (\ref{3.140}), and (\ref{3.147a}),
\begin{equation}
{\rm g}_\sigma(k)=\int_0^k\!dk'\left[\,{1\over b_\sigma}\left(-{c_\sigma\over 2}+b_\sigma\tilde{c}_\sigma-\tilde{b}_\sigma c_\sigma\right)+B\,\right]
\label{3.151}
\end{equation}
with
\begin{equation}
B=B_\sigma+B_{\bar\sigma}
\label{3.148a}
\end{equation}
and
\begin{eqnarray}
\hspace{-5mm}B_\sigma\!\!&=&\!\!-{1\over 4\pi(n_\sigma+n_{\bar\sigma})}\,{k_{T\sigma}\over b_\sigma}\nonumber\\
&\times&\!\!\left\langle {1-\tanh_\sigma\over\zeta^2_\sigma}\left(-{c_\sigma\over 2}+b_\sigma\tilde{c}_\sigma-\tilde{b}_\sigma c_\sigma \right)\right\rangle_\sigma,
\label{3.148b}
\end{eqnarray}
which gives a complete solution to the kinetic equation (\ref{3.40}).
For $\Delta=0$, the constant $B=0$. Recall that $\rho_{\rm D}$ for arbitrary $\Delta$ does not depend on $B$.

\section{Drag resistivity}
\label{s8a}

In Sec.~\ref{s7}, we solved the integro-differential Fokker-Planck equation within the exactly solvable model introduced therein. Apart from providing the analytical expression for the distribution function for an arbitrary relative strength of two- and three-particle scattering, the exact solution also allows us to explicitly and accurately describe the (noncommuting) limits $\Delta\to 0$ and ${\cal D}^{(3)}\to 0$.

In fact, the model solution in Sec.~\ref{s7} is closely related to the solution (not obtainable in the analytical form) of the original Fokker-Planck equation (\ref{3.40}) [with the functions ${\cal D}^{(3)}_\sigma(k)$ and ${\cal C}_\sigma(k)$ resulting from Eqs.~(\ref{3.83d})-(\ref{3.83c})]. Specifically, the exact functions ${\cal D}^{(3)}_\sigma(k)$ and ${\cal C}_\sigma(k)$ can be represented in the form\cite{supplement} that shows that these are slow functions of $k$ compared to the exponentials $\zeta^2_\sigma(k)$. This justifies the approximation of ${\cal D}^{(3)}_\sigma(k)$ and ${\cal C}_\sigma(k)$ by constants, albeit different for different transport regimes, depending on the relative strength of two- and three-particle scattering. In Appendix~\ref{aG}, we also briefly describe how the shape of the function $\partial_k{\rm g}_\sigma(k)$ changes with varying strength of intrawire equilibration (mediated by three-particle scattering) for $\Delta=0$, with a particular goal to justify the approximations that are parametrically accurate for the integral term ${\cal C}_\sigma(k)$. As will be seen in Secs.~\ref{s8} and \ref{s9}, the picture of the evolution of the shape of $\partial_k{\rm g}_\sigma(k)$ from Appendix~\ref{aG} remains valid for arbitrary $\Delta$, with the integral term being important only in the regime of sufficiently strong intrawire equilibration (``drift regime" in the terminology of Ref.~\onlinecite{dmitriev12}), where the representation of ${\cal C}_\sigma$ in terms of the averages $\langle\partial_k{\rm g}_\sigma\rangle$ in Eq.~(\ref{3.116}) is justified parametrically. With this background in mind, we proceed with the calculation of $\rho_{\rm D}$ within the model of Sec.~\ref{s7}.

\subsection{General formula for $\rho_{\rm D}$}
\label{s8b}

Equation~(\ref{3.147}) not only fixes the constant $B$ but, in the form of Eq.~(\ref{3.147b}), together with the relation (\ref{3.146b}), also fixes the matrix structure of $\hat\sigma^{\rm reg}$:
\begin{equation}
\hat\sigma^{\rm reg}=S
\left(\begin{array}{rr}
\!1 & \,-1 \\
\!-1 & \,1 \\
\end{array}\right)~,
\label{3.152}
\end{equation}
where $S$ within the model of Sec.~\ref{s7} is given, as follows from Eq.~(\ref{3.151}), by
\begin{eqnarray}
S&=&{eT\over 4\pi (E_\sigma -E_{\bar\sigma})}\int\!dk\,(1-\tanh_\sigma)\nonumber\\
&\times&\left[\,{1\over b_\sigma}\left(-{c_\sigma\over 2}+b_\sigma\tilde{c}_\sigma-\tilde{b}_\sigma c_\sigma\right)+B\,\right]
\label{3.153a}\\
&=&{eT\over E_\sigma-E_{\bar\sigma}}\left(n_\sigma B_{\bar\sigma}-n_{\bar\sigma}B_\sigma\right)~.
\label{3.153b}
\end{eqnarray}
Note the difference in the matrix structure for $\hat\sigma^{\rm sing}$ [Eq.~(\ref{3.16a})] and $\hat\sigma^{\rm reg}$ [Eq.~(\ref{3.152})]. Substituting Eq.~(\ref{3.152}) in Eq.~(\ref{3.16e}), we obtain the expression for $\rho_{\rm D}$ in terms of $S$:
\begin{equation}
\rho_{\rm D}={n_1n_2\over (n_1+n_2)^2}\,{1\over S}~.
\label{3.150}
\end{equation}
Equation (\ref{3.150}) is general for the relation between $\rho_{\rm D}$ and $S$ [for $S$ understood as the coefficient in front of the matrix in Eq.~(\ref{3.152})], irrespectively of the model to calculate $S$.

\subsection{Ultranarrow resonance at coinciding densities}
\label{s8}

To calculate how the dependence of $\rho_{\rm D}$ on $\Delta$ changes with increasing ${\cal D}^{(3)}$ within the model of Sec.~\ref{s3}, represent $\tilde{b}_\sigma$ and $\tilde{c}_\sigma$ each as a sum of two terms:
\begin{equation}
\tilde{b}_\sigma=\Phi^b_\sigma+\Psi^b_\sigma~,\quad
\tilde{c}_\sigma={e(E_\sigma-E_{\bar\sigma})\over 4(n_\sigma+n_{\bar\sigma})}\left(\Phi^c_\sigma+\Psi^c_\sigma\right)~,
\label{3.154}
\end{equation}
where
\begin{eqnarray}
&&\hspace{-8mm}\Phi^b_\sigma=-{{\cal D}^{(2)}\over 2{\rm D}}\left(\mu_\sigma\zeta^2_\sigma+\lambda_{\bar\sigma}\zeta^2_{\bar\sigma}\right)~,\\
&&\hspace{-8mm}\Psi^b_\sigma=-{{\cal D}^{(3)}_{\bar\sigma}\mu_\sigma\over {\rm D}}~,
\label{3.155a}\\
&&\hspace{-8mm}\Phi^c_\sigma={{\cal D}^{(2)}\over {\rm D}}\left[\,n_{\bar\sigma}(1-\tanh_\sigma)-n_\sigma(1-\tanh_{\bar\sigma})\,\right]~,\\
&&\hspace{-8mm}\Psi^c_\sigma={2{\cal D}^{(3)}_{\bar\sigma} n_{\bar\sigma}\over {\rm D}}\,{1-\tanh_\sigma\over\zeta^2_\sigma}~.
\label{3.155b}
\end{eqnarray}
The term in $\tilde{c}_\sigma$ proportional to $\Phi^c_\sigma$ differs from the remaining part of $\tilde{c}_\sigma$ in that it is singular in ${\cal D}^{(3)}$ for ${\cal D}^{(3)}\to 0$. More specifically, $\Phi^c_\sigma$ is singular as ${\cal O}[\Delta/{\cal D}^{(3)}]$ for ${\cal D}^{(3)}\to 0$ and $\Delta\to 0$, so that its limiting value depends on the order of taking the two limits. The term $\Phi^b_\sigma$ differs from $\Psi^b_\sigma$ in that it is finite for ${\cal D}^{(3)}\to 0$, whereas $\Psi^b_\sigma$ vanishes in this limit. For $\Delta=0$, the term $\Phi^c_\sigma$ vanishes for any ${\cal D}^{(3)}\neq 0$, and $S$ in the limit of ${\cal D}^{(3)}\to 0$ is finite (not infinite) and given by the contribution of two-particle scattering. For $\Delta\neq 0$, the term $\Phi^c_\sigma$ gives a contribution to $S$ which diverges as $1/{\cal D}^{(3)}$ for ${\cal D}^{(3)}\to 0$. This divergence means $\hat\rho (\omega=0)=0$ for $\Delta\neq 0$, i.e., $\rho_{\rm D}$ as a function of $\Delta$ is a peak of zero width if only two-particle scattering is present.

In terms of $\Phi^{b,c}_\sigma$ and $\Psi^{b,c}_\sigma$, $S$ is rewritten as
\begin{eqnarray}
&&\hspace{-3mm}S={e^2T\over 16\pi (n_\sigma+n_{\bar\sigma})^2}\left\{n_{\bar\sigma}\!\int\!dk\,(1-\tanh_\sigma)\right.\nonumber\\
&&\hspace{-3mm}\times\left[-\left(\left\langle\Phi_\sigma^c\right\rangle_\sigma+\left\langle\Psi_\sigma^c
\right\rangle_\sigma\right){1/2+\Phi^b_\sigma+\Psi^b_\sigma\over \left\langle\Phi^b_\sigma\right\rangle_\sigma+\left\langle\Psi^b_\sigma\right\rangle_\sigma}+\Phi^c_\sigma+\Psi^c_\sigma\right]\nonumber\\
&&\hspace{30mm}+\,(\sigma\leftrightarrow\bar\sigma)\bigg\}~.
\label{3.156}
\end{eqnarray}
For $\Delta=0$ ($\Phi^c_\sigma=0$), Eq.~(\ref{3.156}) is identical to $1/4\rho_{\rm D}$ from Ref.~\onlinecite{dmitriev12}, with $1/\rho_{\rm D1}$ [Eq.~(3.66) in Ref.~\onlinecite{dmitriev12}] associated with the term $\Psi^c_\sigma$ [i.e., with the last term in the square brackets of Eq.~(\ref{3.156}), but not with the average $\langle\Psi^c_\sigma\rangle_\sigma$ in the round brackets] and $1/\rho_{\rm D2}$ [Eq.~(3.67) in Ref.~\onlinecite{dmitriev12}] with the rest.

Let us first calculate $\rho_{\rm D}$ in the tail of the peak, broadened by three-particle scattering, in the limit of small ${\cal D}^{(3)}$ (``weak intrawire equilibration"). To this end, we retain in $S$ only the terms $\Phi^{b,c}_\sigma$ by dropping the terms $\Psi^{b,c}_\sigma$; moreover, in the denominator of $\Phi^{b,c}_\sigma$, we retain only the term in $\rm D$ that is linear in ${\cal D}^{(3)}_\sigma$, i.e., drop the term ${\cal D}^{(3)}_\sigma{\cal D}^{(3)}_{\bar\sigma}$. In this limit [${\cal D}^{(3)}\to 0$ for $\Delta\neq 0$], $S$ becomes independent of ${\cal D}^{(2)}$. For $|\Delta|\ll T$ [recall that the peak becomes bodyless in the limit of small ${\cal D}^{(3)}$], it suffices to expand $S$ to second order in $\Delta$. For $|\Delta|\ll \epsilon_{F\sigma}$, one can also neglect the dependence of ${\cal D}^{(3)}_\sigma$ on $\sigma$, so that below we express $S$ in terms of $\lambda$, $\mu$, and ${\cal D}^{(3)}=\lambda+\mu$, by dropping the index $\sigma$ and writing $\Phi_\sigma^{b,c}$ as
\begin{eqnarray}
&&\hspace{-5mm}\Phi^b_\sigma\simeq M^b_\sigma=-{1\over {\cal D}^{(3)}}\,{\mu\zeta^2_\sigma+\lambda\zeta^2_{\bar\sigma}\over\zeta^2_\sigma +\zeta^2_{\bar\sigma}}~,\label{3.157a}\\
&&\hspace{-5mm}\Phi^c_\sigma\simeq {2\over{\cal D}^{(3)}}M^c_\sigma~,\nonumber\\
&&\hspace{-5mm}M^c_\sigma={n_{\bar\sigma}(1-\tanh_\sigma)-n_\sigma (1-\tanh_{\bar\sigma})\over \zeta^2_\sigma +\zeta^2_{\bar\sigma}}~.
\label{3.157b}
\end{eqnarray}
To find $S$ to order ${\cal O}[\Delta^2/{\cal D}^{(3)}]$, we need to expand $M^b_\sigma$ and $M^c_\sigma$ to first and second order in $\Delta$, respectively,
\begin{eqnarray}
&&M^b_\sigma=-{1\over 2}+w_\sigma +{\cal O}(\Delta^2)~,\\
&&M^c_\sigma=u_\sigma + v_\sigma +{\cal O}(\Delta^3)~,
\label{3.159}
\end{eqnarray}
where
\begin{eqnarray}
&&\hspace{-8mm}u_\sigma={1\over 2}(\epsilon_{F\bar\sigma}-\epsilon_{F\sigma})\left[\,\left({\partial n\over\partial\mu}\right)_\sigma \!{1-\tanh_\sigma\over \zeta^2_\sigma} -{n_\sigma\over 2T}\right]\nonumber\\
&&\hspace{46mm}\sim {\cal O}(\Delta)~,
\label{3.160a}\\
&&\hspace{-8mm}v_\sigma=-(\epsilon_{F\bar\sigma}-\epsilon_{F\sigma})^2\,\left({\partial n\over\partial\mu}\right)_\sigma {1\over 4T}{(1-\tanh_\sigma)\tanh_\sigma\over \zeta^2_\sigma}\nonumber\\
&&\hspace{46mm}\sim {\cal O}(\Delta^2)~,
\label{3.160b}\\
&&\hspace{-8mm}w_\sigma=-(\epsilon_{F\bar\sigma}-\epsilon_{F\sigma})\,{\lambda-\mu\over{\cal D}^{(3)}}{1\over 4T}\tanh_\sigma\sim {\cal O}(\Delta)~,
\label{3.160c}
\end{eqnarray}
and $(\partial n/\partial\mu)_\sigma$, the compressibility of the electron gas, is given by Eq.~(\ref{3.116d}).

To order ${\cal O}[\Delta^2/{\cal D}^{(3)}]$ for $S$, we have
\begin{eqnarray}
S\!\!&\simeq&\!\!{e^2T\over 32\pi n^2}\,{1\over {\cal D}^{(3)}}\left[\,n_{\bar\sigma}\!\int\!dk\,(1-\tanh_\sigma)\right.\nonumber\\
&\times&\!\!\left(u_\sigma +v_\sigma +2\left\langle u_\sigma\right\rangle_\sigma w_\sigma\right)+(\sigma\leftrightarrow\bar\sigma)\,\bigg]~,
\label{3.161}
\end{eqnarray}
where $n$ is the density in the wires at $\Delta=0$. Since $n_\sigma$ (equal to $k_{F\sigma}/\pi$ at $T=0$) is given for arbitrary $T$ by Eq.~(\ref{4.20}), which can be rewritten as
\begin{equation}
n_\sigma={1\over 4\pi}\!\int\!dk\,(1-\tanh_\sigma)~,
\label{3.161a}
\end{equation}
the average $\langle u_\sigma\rangle_\sigma$ vanishes exactly,
\begin{equation}
\langle u_\sigma\rangle_\sigma=0~,
\label{3.162}
\end{equation}
so that the term $w_\sigma$ drops out from Eq.~(\ref{3.161}). Note that Eq.~(\ref{3.162}) results from a cancelation of two integrals, corresponding to two terms in the square brackets of Eq.~(\ref{3.160a}), determined by vastly different momenta: one determined by all $k<k_{F\sigma}$ and the other determined by $|k|-k_{F\sigma}\sim T/v_{F\sigma}$. Note also that, since $\Phi^c_\sigma=-\Phi^c_{\bar\sigma}$ is antisymmetric in $\sigma$, the vanishing of $\langle u_\sigma\rangle_\sigma$ is required by the relation (\ref{3.139}) from which $\left\langle\Phi^c_\sigma\right\rangle_\sigma\to\left\langle\Phi^c_{\bar\sigma}\right\rangle_{\bar\sigma}$ for $\Delta\to 0$ at order ${\cal O}[1/{\cal D}^{(3)}]$ for ${\cal D}^{(3)}\to 0$. To ensure the cancelation (\ref{3.162}), it is important to use the exact ($T$ dependent) compressibility $(\partial n/\partial\mu)_\sigma$ [Eq.~(\ref{3.116d})] in Eq.~(\ref{3.160a}), i.e., not to substitute $1/\pi v_{F\sigma}$ for it. The term $v_\sigma$ should be retained in Eq.~(\ref{3.161}), written in terms of both $u_1$ and $u_2$, at order ${\cal O}(\Delta^2)$; however, from the relation $M^c_\sigma=-M^c_{\bar\sigma}$, Eq.~(\ref{3.161}) can be rewritten, after the cancelation (\ref{3.162}), as
\begin{eqnarray}
&&\hspace{-12mm}S\simeq {e^2T\over 32\pi n^2}\,{1\over {\cal D}^{(3)}}\nonumber\\
&&\hspace{-9mm}\times\!\int\!dk\,\left[\,n_{\bar\sigma}(1-\tanh_\sigma)-n_\sigma (1-\tanh_{\bar\sigma})\,\right] u_\sigma~,
\label{3.163}
\end{eqnarray}
with the term $v_\sigma$ in the sum $u_\sigma+v_\sigma$ being in Eq.~(\ref{3.163}) beyond the accuracy at order ${\cal O}(\Delta^2)$ for $S$. Substituting Eq.~(\ref{3.160a}) in Eq.~(\ref{3.163}), we obtain, at order ${\cal O}[\Delta^2/{\cal D}^{(3)}]$ (and arbitrary $T/\epsilon_F$):
\begin{eqnarray}
S\!\!&\simeq&\!\! {e^2T\over 64\pi n^2}\,{\Delta^2\over {\cal D}^{(3)}}\,{\partial n\over\partial\mu}\,(2\pi mT)^{1/2}\nonumber\\
&\times&\!\!\left[\,{\partial n\over\partial\mu}\,e^{\epsilon_F/T}-n^2\left({2\pi\over mT^3}\right)^{1/2}\,\right]
\label{3.164}
\end{eqnarray}
(where we dropped the index $\sigma$ everywhere), which for $T\ll\epsilon_F$ reduces to
\begin{equation}
S\simeq {e^2\over 128}\left({mT^3\over 2\pi}\right)^{1/2}\!\left({\Delta\over\epsilon_F}\right)^2{1\over {\cal D}^{(3)}}\,e^{\epsilon_F/T}~.
\label{3.165}
\end{equation}

To obtain the broadening of the peak in $\rho_{\rm D}$ as a function of $\Delta$ in the limit of small ${\cal D}^{(3)}$, we add to $S$ from Eq.~(\ref{3.165}), calculated at order ${\cal O}[\Delta^2/{\cal D}^{(3)}]$, the contribution $S^{(2)}$ of two-particle scattering for identical wires, obtainable by putting $\lambda=\mu=0$ in Eqs.~(\ref{3.54b}) and (\ref{3.54c}) and substituting the result in Eqs.~(\ref{3.148b}) and (\ref{3.153b}), at $\Delta=0$ (see also Appendix~\ref{aD}). For $T\ll\epsilon_F$, $S^{(2)}$ reads
\begin{equation}
S^{(2)}\simeq {e^2\over 64}\left({mT^3\over \pi}\right)^{1/2}\!{1\over {\cal D}^{(2)}}\,e^{2\epsilon_F/T}~.
\label{3.166}
\end{equation}
For the dependence of $\rho_{\rm D}$ on $\Delta$ we thus obtain a Lorentzian (Fig.~\ref{f1}):
\begin{equation}
\rho_{\rm D}(\Delta)\simeq\rho_{\rm D}(0)\,{1\over 1+(\Delta/\Gamma)^2}~,
\label{3.167}
\end{equation}
where
\begin{equation}
\rho_{\rm D}(0)={16{\cal D}^{(2)}\over e^2nT}\left({2\epsilon_F\over \pi T}\right)^{1/2}e^{-2\epsilon_F/T}
\label{3.168}
\end{equation}
and
\begin{equation}
\Gamma=\epsilon_F\left[\,{2^{1/2}{\cal D}^{(3)}\over {\cal D}^{(2)}}e^{\epsilon_F/T}\right]^{1/2}~.
\label{3.169}
\end{equation}
As a function of the density mismatch $\Delta_n=n_1-n_2$, given by $2m\Delta/\pi^2(n_1+n_2)$ for $T=0$, with $\Delta_n=\Delta/\pi v_F$ in the limit of small $\Delta$, the drag resistivity is a Lorentzian $\rho_{\rm D}(\Delta_n)\simeq\rho_{\rm D}(0)/[1+(\Delta_n/\Gamma_n)^2]$ with
\begin{equation}
\Gamma_n= n\left[\,{{\cal D}^{(3)}\over 2^{1/2}{\cal D}^{(2)}}e^{\epsilon_F/T}\right]^{1/2}~.
\label{3.170}
\end{equation}

The conditions of applicability for the derivation of Eq.~(\ref{3.165}) for $S$ at order ${\cal O}[\Delta^2/{\cal D}^{(3)}]$ were: (i) $|\Delta|\ll T$, (ii) $\Gamma\ll |\Delta|$, and (iii) ${\cal D}^{(3)}\ll{\cal D}^{(2)}e^{-\epsilon_F/T}$. The first condition was used in the expansion in powers of $\Delta$ in Eqs.~(\ref{3.160a})-(\ref{3.160c}). The second and third ones were used to justify the neglect of $\Psi^{b,c}$ compared to $\Phi^{b,c}$ and the omitting of the term ${\cal D}^{(3)}_\sigma{\cal D}^{(3)}_{\bar\sigma}$ in ${\rm D}$, respectively. For the precise form of the third condition, it is important that the first term in the square brackets in Eq.~(\ref{3.164}) (the main one for $T\ll\epsilon_F$) is determined by momenta $|k|\alt (mT)^{1/2}$ in the integral (\ref{3.163}). For the body of the peak of $\rho_{\rm D}(\Delta)$ to be described by Eqs.~(\ref{3.167})-(\ref{3.169}), the condition $\Gamma\ll T$ must be fulfilled, i.e.,
\begin{equation}
{\cal D}^{(3)}/{\cal D}^{(2)}\ll (T/\epsilon_F)^2e^{-\epsilon_F/T}~,
\label{3.171}
\end{equation}
which means that the condition $\Gamma\ll T$ for $T\ll\epsilon_F$ is more stringent than ${\cal D}^{(3)}\ll{\cal D}^{(2)}e^{-\epsilon_F/T}$. We conclude that the broadening of the peak of $\rho_{\rm D}$ as ${\cal D}^{(3)}$ increases is described by Eqs.~(\ref{3.167})-(\ref{3.169}) for ${\cal D}^{(3)}$ satisfying Eq.~(\ref{3.171}), i.e., if this condition is fulfilled, the shape of the peak is different from the Lorentzian only far in the tails for $|\Delta|\agt T$. Note that, while the peak in Eq.~(\ref{3.167}) becomes broader with increasing ${\cal D}^{(3)}$ for given ${\cal D}^{(2)}$, its amplitude remains almost constant.

\subsection{Strong intrawire equilibration}
\label{s9}

Equilibration processes mediated by triple collisions are solely responsible for the broadening of the ultranarrow resonance in Eqs.~(\ref{3.167})-(\ref{3.169}), with the resonance width $\Gamma\propto [{\cal D}^{(3)}]^{1/2}$ (Fig.~\ref{f1}). When $\Gamma$ becomes, with increasing ${\cal D}^{(3)}$, of the order of $T$, the dissipation processes that give a nonzero $\rho_{\rm D}$ at $\Delta\neq 0$ are no longer ``bottlenecked" by triple collisions. Let us now turn to this regime of a strong (in the sense specified above) equilibration induced by triple collisions. For
${\cal D}^{(3)}/{\cal D}^{(2)}\gg (T/\epsilon_F)^2e^{-\epsilon_F/T}$ [the condition opposite to Eq.~(\ref{3.171})], the term $\Phi^c_\sigma$ can be neglected in $S$. Specifically, for ${\cal D}^{(3)}/{\cal D}^{(2)}\ll e^{\epsilon_F/T}(T/\epsilon_F)^{3/2}$ [this condition means\cite{dmitriev12} that triple collisions are not yet capable of establishing the fully equilibrated drift regime], the main contribution to $S$ is given by $\Psi^c_\sigma$, the last term in the square brackets in Eq.~(\ref{3.156}). For $|\Delta|\ll \epsilon_F$, we have
\begin{equation}
S\simeq {e^2T\over 64\pi n}\int\!dk\,\left[(1-\tanh_\sigma)\Psi^c_\sigma + (1-\tanh_{\bar\sigma})\Psi^c_{\bar\sigma}\right]~.
\label{3.172}
\end{equation}
For ${\cal D}^{(3)}/{\cal D}^{(2)}\ll e^{-\epsilon_F/T}$, the term in $\rm D$ quadratic in ${\cal D}^{(3)}$ can still be neglected, and $\Psi^c_\sigma$ can be written as
\begin{equation}
\Psi^c_\sigma\simeq {4n\over {\cal D}^{(2)}}{1-\tanh_\sigma\over\zeta^2_\sigma (\zeta^2_\sigma+\zeta^2_{\bar\sigma})}~.
\label{3.173}
\end{equation}
More precisely, $\Psi^c_\sigma$ is given by Eq.~(\ref{3.173}) in the limit of small ${\cal D}^{(3)}$ for not too large $|k|$ above $k_F$, namely for $|k|<k_F\ln^{1/2}[{\cal D}^{(2)}/{\cal D}^{(3)}]$. Similar to Ref.~\onlinecite{dmitriev12}, we assume that the contribution of $|k|\gg k_F$ to $S$ for ${\cal D}^{(3)}\to 0$, which is cut off by higher-order gradient terms in the collision integral that were neglected in the Fokker-Planck approximation, is smaller than the contribution of $k$ on and below the Fermi surface. The integral (\ref{3.172}) is then determined by $|k|\sim (mT)^{1/2}$, so that $1-\tanh_\sigma$ can be substituted with 2 in Eqs.~(\ref{3.172}) and (\ref{3.173}), with
\begin{equation}
\Psi^c_\sigma+\Psi^c_{\bar\sigma}\simeq {8n\over {\cal D}^{(2)}}{1\over\zeta^2_\sigma\zeta^2_{\bar\sigma}}~,
\label{3.173a}
\end{equation}
and we obtain
\begin{equation}
\rho_{\rm D}\simeq {16{\cal D}^{(2)}\over e^2}\left({\pi\over mT^3}\right)^{1/2}e^{-(\epsilon_{F1}+\epsilon_{F2})/T}~.
\label{3.174}
\end{equation}

\begin{figure}
\centerline{\includegraphics[width=\columnwidth]{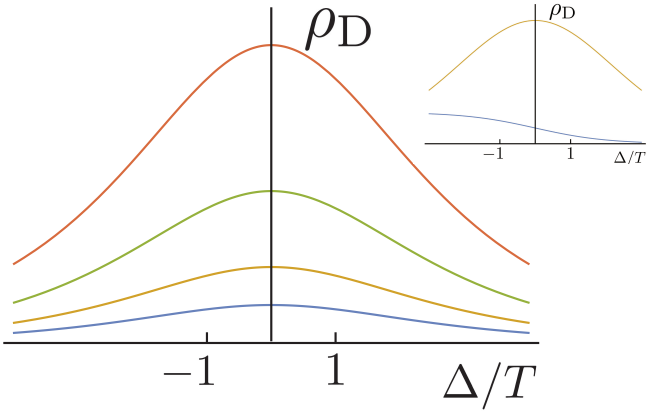}}
\caption{Schematic dependence of the drag resistivity $\rho_{\rm D}$ for two quantum wires on the chemical potential mismatch $\Delta$ (in units of the temperature $T$) for the case of strong intrawire equilibration [such that the ultranarrow resonance peak with the width $\Gamma\ll T$ (Fig.~\ref{f1}) is completely washed out]. The curves differ in the strength of three-particle scattering (for fixed strength of two-particle scattering) which increases from bottom to top [with the shape of the curves varying between those given by Eq.~(\ref{3.176}) for the lower curve and Eq.~(\ref{3.184}) for the upper]. The curves in the main part of the figure are for a symmetric splitting of the Fermi levels (with $\epsilon_{F1}+\epsilon_{F2}$ held constant). Those in the inset are for an asymmetric splitting with $\epsilon_{F2}$ kept fixed. The lower and upper curves in the inset correspond to Eqs.~(\ref{3.176}) and (\ref{3.184}), respectively.}
\label{f2}
\end{figure}

Note that $\rho_{\rm D}$ in Eq.~(\ref{3.174}) does not depend on ${\cal D}^{(3)}$, although the drag regime described by Eq.~(\ref{3.174}) is only possible for $\Delta\neq 0$ because of sufficiently strong three-particle scattering. The activation gap in Eq.~(\ref{3.174}) is given by the sum of the Fermi energies, so that $\rho_{\rm D}$ can both increase and decrease with varying asymmetry, depending on precisely in what way the wires are unbalanced. In particular, if one of the Fermi energies is increased and the other decreased in such a way that their sum remains constant, $\rho_{\rm D}$ in the limit $|\Delta|\ll\epsilon_{F\sigma}$ (but for arbitrary $|\Delta|/T$) stays constant as well. If, however, one of the Fermi energies increases or decreases while the other remains unchanged, $\rho_{\rm D}$ decreases or increases, respectively.

Using Eq.~(\ref{3.172}) in the plateau regime [see Eq.~(3.68) in Ref.~\onlinecite{dmitriev12}], for $e^{-\epsilon_F/T}\ll {\cal D}^{(3)}/{\cal D}^{(2)}\ll e^{\epsilon_F/T}(T/\epsilon_F)^{3/2}$, with
\begin{equation}
\Psi^c_\sigma\simeq {2n\over {\cal D}^{(3)}}{1-\tanh_\sigma\over\zeta^2_\sigma}~,
\label{3.175}
\end{equation}
yields
\begin{eqnarray}
\rho_{\rm D}\!\!&\simeq&\!\! {8{\cal D}^{(3)}\over e^2}\left({\pi\over 2mT^3}\right)^{1/2}{1\over e^{\epsilon_{F1}/T}+e^{\epsilon_{F2}/T}}
\nonumber\\
&=&\!\!{4{\cal D}^{(3)}\over e^2}\left({\pi\over 2mT^3}\right)^{1/2}{e^{-(\epsilon_{F1}+\epsilon_{F2})/2T}\over \cosh (\Delta/2T)}~.
\label{3.176}
\end{eqnarray}
The strength of three-particle scattering ${\cal D}^{(3)}$ ``reemerges" in the expression (\ref{3.176}) for $\rho_{\rm D}$ [cf.\ Eqs.~(\ref{3.167})-(\ref{3.169}) and (\ref{3.174})]. Similar to Eq.~(\ref{3.174}), the integral (\ref{3.172}) with $\Psi^c_\sigma$ from Eq.~(\ref{3.175}) is determined by $|k|\sim (mT)^{1/2}$. In contrast to Eq.~(\ref{3.174}), however, $\rho_{\rm D}$ changes (decreases) if the Fermi levels are split in a symmetric way (with $\epsilon_{F1}+\epsilon_{F2}$ being constant), see Fig.~\ref{f2}. Moreover, the gap in the Arrhenius law in Eq.~(\ref{3.176}) is given by $\max\{\epsilon_{F1},\epsilon_{F2}\}$, so that the gap does not change if one of the Fermi energies decreases (inset in Fig.~\ref{f2}), which is also in contrast to Eq.~(\ref{3.174}).

For ${\cal D}^{(3)}/{\cal D}^{(2)}\gg e^{\epsilon_F/T}(T/\epsilon_F)^{3/2}$, the main contribution to $S$ comes from the term in Eq.~(\ref{3.156}) proportional to $\langle\Psi^c_\sigma\rangle_\sigma$:
\begin{eqnarray}
&&\hspace{-10mm}S\simeq -{e^2T\over 64\pi n}\nonumber\\
&&\hspace{-10mm}\times\int\!dk\left[\,(1-\tanh_\sigma)\,{\langle\Psi^c_\sigma\rangle_\sigma \Theta_\sigma\over\langle\Theta_\sigma\rangle_\sigma -1/2}
+(\sigma\leftrightarrow{\bar\sigma})\,\right],
\label{3.177}
\end{eqnarray}
where
\begin{equation}
\Theta_\sigma={1\over 2}+\Phi^b_\sigma+\Psi^b_\sigma
={\lambda-\mu\over 2{\cal D}^{(3)}}\,{{\cal D}^{(2)}(\zeta^2_\sigma-\zeta^2_{\bar\sigma})+2{\cal D}^{(3)}\over {\cal D}^{(2)}(\zeta^2_\sigma+\zeta^2_{\bar\sigma})+2{\cal D}^{(3)}}
\label{3.181a}
\end{equation}
with
\begin{eqnarray}
&&\hspace{-5mm}\int\!dk\,(1-\tanh_\sigma)\Theta_\sigma\simeq 2\pi n\,{\lambda-\mu\over {\cal D}^{(3)}}~,\label{3.180a}\\
&&\hspace{-5mm}\langle\Theta_\sigma\rangle_\sigma\simeq {\lambda-\mu\over 2{\cal D}^{(3)}}\left[\,1-{{\cal D}^{(2)}\over {\cal D}^{(3)}}\,{v_F\over 8T}\!\int\!dk\,\zeta^2_\sigma\zeta^2_{\bar\sigma}\,\right],
\label{3.180b}
\end{eqnarray}
and
\begin{equation}
\langle\Psi^c_\sigma\rangle_\sigma={4 n_\sigma\over k_{T\sigma}}\int\!dk\,{1-\tanh_\sigma\over {\cal D}^{(2)}(\zeta^2_\sigma+\zeta^2_{\bar\sigma})+2{\cal D}^{(3)}}\simeq {2n\over{\cal D}^{(3)}}\,{\epsilon_F\over T}~.
\label{3.181}
\end{equation}
The integral in Eq.~(\ref{3.180b}) is given for $T,|\Delta|\ll\epsilon_F$ by
\begin{equation}
{v_F\over 8T}\int\!dk\,\zeta^2_\sigma\zeta^2_{\bar\sigma}\simeq {\cal J}\!\left({\Delta\over 2T}\right)~,
\label{3.182}
\end{equation}
where
\begin{equation}
{\cal J}(x)={2\over\sinh^2 x}\left(x\coth x-1\right)~.
\label{3.183}
\end{equation}
Note that the integrals in the numerator of $S$, namely those in Eqs.~(\ref{3.180a}) and (\ref{3.181}), are determined by all $k$ below the Fermi surface(s), whereas the integral in the denominator [Eq.~(\ref{3.180b})] is determined by $k$ in the vicinity of the Fermi surface(s).

For ${\cal D}^{(3)}/{\cal D}^{(2)}\gg e^{\epsilon_F/T}(T/\epsilon_F)^{3/2}$,
we thus obtain from Eq.~(\ref{3.177}):
\begin{equation}
\rho_{\rm D}\simeq {1\over e^2n\epsilon_F}\left[\,{\cal D}^{(2)}{\cal J}\!\left({\Delta\over 2T}\right)+2\mu\,\right]
\label{3.184}
\end{equation}
[where we also used ${\cal D}^{(3)}=\lambda+\mu\gg\mu$, i.e., three-particle scattering inside each of the wires is assumed to be much stronger than that between the wires, which is generically consistent with the condition ${\cal D}^{(3)}\gg {\cal D}^{(2)}$]. The first and second terms in Eq.~(\ref{3.184}) can be viewed as the contributions of two- and three-particle scattering to $\rho_{\rm D}$, respectively. However, it is important to emphasize that the contribution of two-particle scattering has this form, corresponding to the drift regime, only for sufficiently strong three-particle scattering\cite{remark4} (specifically, for sufficiently strong equilibration of electrons in each of the wires among themselves).

For $\Delta=0$, Eq.~(\ref{3.184}) [with ${\cal J}(0)=2/3$] reproduces the result of Ref.~\onlinecite{dmitriev12} for the drift regime, with the two-particle term giving the main contribution to $\rho_{\rm D}$. Splitting of the Fermi levels is thus seen to strongly suppress drag in Eq.~(\ref{3.184}) by reducing (exponentially in $|\Delta|/T$) the contribution of two-particle scattering (Fig.~\ref{f2}), in agreement with the results of Refs.~\onlinecite{pustilnik03} and \onlinecite{aristov07}. As $|\Delta|$ increases, the exponential suppression of $\rho_{\rm D}$ saturates at the contribution of three-particle scattering
\begin{equation}
\rho_{\rm D}\simeq 2\mu/e^2n\epsilon_F~.
\label{3.185}
\end{equation}

A similar behavior for $\rho_{\rm D}(\Delta)$ in the drift regime was suggested in Refs.~\onlinecite{pustilnik03} and \onlinecite{aristov07} (with a parametrically larger $\rho_{\rm D}$ in the latter, see Ref.~\onlinecite{aristov07} for more detail) as a result of two-particle interactions inside the wires being taken into account alongside with two-particle interactions between the wires. More specifically, the saturation of the exponential falloff of $\rho_{\rm D}(\Delta)$ was associated in Refs.~\onlinecite{pustilnik03} and \onlinecite{aristov07} with the existence of a power-law tail of the dynamic structure factor for a given wire, produced by intrawire interactions, leading to a power-law (in $\Delta$) overlap between the dynamic structure factors for two wires for sufficiently large $|\Delta|$. The resulting contribution to $\rho_{\rm D}$, which avoids the exponential suppression with increasing $|\Delta|/T$, is of fourth order in interaction $\sim {\cal O}(V_{\sigma\sigma}^2V_{\sigma\bar\sigma}^2)$ (two powers of the intrawire interaction $V_{\sigma\sigma}$ and two powers of the interwire interaction $V_{\sigma\bar\sigma}$). As was pointed out in Ref.~\onlinecite{aristov07}, a similar contribution to $\rho_{\rm D}$ exists also at order ${\cal O}(V_{\sigma\bar\sigma}^4)$. The fourth order in interaction, with at least two powers of the interwire interaction, is precisely the order at which three-particle scattering contributes\cite{supplement} to $\mu$ in Eq.~(\ref{3.185}). There is, however, an important difference: when solving the kinetic equation, we obtain $\rho_{\rm D}$ of fourth order in interaction in Eq.~(\ref{3.185}) as associated with three-particle scattering, rather than with independent two-particle scattering events renormalized (``dressed") by virtual processes in Refs.~\onlinecite{pustilnik03} and \onlinecite{aristov07}.

\section{Summary}
\label{s10}

We have discussed Coulomb drag between nonidentical ballistic quantum wires within the kinetic-equation formalism. The conventional theory of Coulomb drag, implicitly presupposing infinitely fast intrawire equilibration, has proven to be totally inadequate to describe the behavior of the dc drag resistivity $\rho_{\rm D}$ as a function of the chemical potential difference between the wires $\Delta$. One ``unexpected" feature of Coulomb drag in one dimension that we have demonstrated in this paper is the exact vanishing of $\rho_{\rm D}$ at any nonzero $\Delta$, for Coulomb drag mediated by two-particle scattering, even though $\rho_{\rm D}$ is finite at $\Delta=0$ [Eq.~(\ref{4.24})]. Further, we have shown that the resonance in $\rho_{\rm D}$ at $\Delta=0$ is broadened by processes of three-particle scattering [Eq.~(\ref{3.167})], which emphasizes the importance of multi-particle scattering for transport in one dimension (for a typical experimental situation in the Coulomb drag problem, the resonance is likely to be washed out by triple collisions). We have also calculated $\rho_{\rm D}(\Delta)$ for the Coulomb drag regimes in which the resonance at $\Delta=0$ is completely destroyed by three-particle scattering. In a wide range of the parameters of the problem, $\rho_{\rm D}$ shows then an activation behavior with the gap given by the largest between the Fermi energies in two wires [Eq.~(\ref{3.176})].

\acknowledgments

This work was supported by the Russian Foundation for Basic Research under Grant No.\ 15-02-04496-a.

\begin{appendix}

\vspace{4mm}

\begin{widetext}

\vspace{20mm}

\section{Fokker-Planck collision integral}
\label{aB}
\renewcommand{\theequation}{A.\arabic{equation}}
\setcounter{equation}{0}

The current $J_\sigma$ [Eq.~(\ref{3.12})], corresponding to the collision integrals (\ref{4.3}) and (\ref{3.3a})-(\ref{3.3c}), is a sum of two- [$J_\sigma^{(2)}$] and three-particle contributions [$J_\sigma^{(3a),(3b),(3c)}$]. In the Fokker-Planck limit, i.e., for $T\gg v_{F\sigma}/a$, these are given by
\begin{eqnarray}
J_\sigma^{(2)}(k)&\simeq& D^{(2)}(k)\,{1\over 2}\partial_k[\,g_\sigma(k)-g_{\bar\sigma}(k)\,]~,\label{3.21a}\\
J_\sigma^{(3a),(3b),(3c)}(k)&\simeq& D_\sigma^{(3a),(3b),(3c)}(k)\,\partial_kg_\sigma(k)-C_\sigma^{(3a),(3b),(3c)}(k)~,
\label{3.21b}
\end{eqnarray}
where
$D^{(2)}(k)={\cal D}^{(2)}\zeta^2_\sigma(k)\zeta^2_{\bar\sigma}(k)$
[with ${\cal D}^{(2)}$ from Eq.~(\ref{4.12})],
\begin{eqnarray}
D_\sigma^{(3a),(3b),(3c)}(k)&=&{1\over 2}\xi_{a,b,c}\int\!\!\!\!\!\!-\,{dq\over 2\pi}\,q^2{\cal P}^{(3a),(3b),(3c)}_\sigma(k,k+q)~,
\label{3.23a}\\
C_\sigma^{(3a),(3b),(3c)}(k)&=&\xi_{a,b,c}\int\!\!\!\!\!\!-\,{dq\over 2\pi}\,q\,\bar{\cal P}^{(3a),(3b),(3c)}_\sigma(k,k+q)
\label{3.23b}
\end{eqnarray}
with $\xi_{a,b,c}=3\,,1\,,2$, and
\begin{eqnarray}
{\cal P}^{(3a),(3b),(3c)}_\sigma(1,1')&=&{1\over 4}\eta_{a,b,c}\,L\!\sum_{232'3'}W_\sigma^{(3a),(3b),(3c)}(1',2',3'|1,2,3)\delta_3(\ldots)~,
\label{3.18c}\\
\bar{\cal P}^{(3a)}_\sigma(1,1')&=&{1\over 4}\eta_a\,L\!\sum_{232'3'}W_\sigma^{(3a)}(1',2',3'|1,2,3)\,\delta_3(\ldots)\,[\,g_\sigma(2)-g_\sigma(2')\,]~,
\label{3.18d}\\
\bar{\cal P}^{(3b)}_\sigma(1,1')&=&{1\over 4}\eta_b\,L\!\sum_{232'3'}W_\sigma^{(3b)}(1',2',3'|1,2,3)\,\delta_3(\ldots)\,[\,g_{\bar\sigma}(2)-g_{\bar\sigma}(2')\,]~,
\label{3.18e}\\
\bar{\cal P}^{(3c)}_\sigma(1,1')&=&{1\over 4}\eta_{c}\,L\!\sum_{232'3'}W_\sigma^{(3c)}(1',2',3'|1,2,3)\,\delta_3(\ldots)
\,{1\over 2}\,[\,g_\sigma(2)-g_\sigma(2')+g_{\bar\sigma}(3)-g_{\bar\sigma}(3')\,]~.
\label{3.18f}
\end{eqnarray}
If $\Delta\neq 0$, the diffusion coefficients $D_\sigma^{(3a),(3b),(3c)}(k)$ and the integral terms $C_\sigma^{(3a),(3b),(3c)}(k)$ in the current induced by three-particle scattering depend on $\sigma$, whereas the diffusion coefficient $D^{(2)}(k)$ for two-particle scattering does not.
The dash in $\int\hspace{-3.1mm}-$ in Eqs.~(\ref{3.23a}) and (\ref{3.23b}) means that the integration is understood as performed only over $|q|\alt 1/a\ll T/v_{F1},T/v_{F2}$, with the contribution of exchange processes with $|q|\gg 1/a$ being accounted for by the factors $\xi_{a,b,c}$ (in Ref.~\onlinecite{dmitriev12}, this constraint was implicitly understood but not marked in the formulas).

As a function of $q$ for given $k$, ${\cal P}^{(3a)}_\sigma(k,k+q)$ shows three peaks centered at $q=0$ and $\pm k_{F\sigma}-k$. The characteristic width of the peak at $q=0$ is $1/a$, whereas that of two other peaks is $T/v_{F\sigma}\gg 1/a$. In the Fokker-Planck limit, the weight of all three peaks is the same, hence $\xi_a=3$. In channel (c), too, there are three peaks in ${\cal P}^{(3c)}_\sigma(k,k+q)$, centered at the same points; however, the weight of each of the peaks around $q=\pm k_{F\sigma}-k$ is half that of the peak at $q=0$, i.e., $\xi_c=2$. In channel (b), there is a single peak at $q=0$, so that $\xi_b=1$. The separation of the contribution of the sharp peak at $q=0$ from that of the broad exchange-induced [in channels (a) and (c)] ``satellite" peaks is parametrically accurate in the Fokker-Planck limit, independently of $k$ (including the case of $||k|-k_{F\sigma}|\alt T/v_{F\sigma}$, where the peak at $q=0$ overlaps with one of the peaks at $q=\pm k-k_{F\sigma}$).

In Eq.~(\ref{3.23b}), the differences $g_\sigma(2)-g_\sigma(2')$ in $\bar{\cal P}^{(3a)}_\sigma$, $g_{\bar\sigma}(2)-g_{\bar\sigma}(2')$ in $\bar{\cal P}^{(3b)}_\sigma$, and $g_\sigma(2)-g_\sigma(2')+g_{\bar\sigma}(3)-g_{\bar\sigma}(3')$ in $\bar{\cal P}^{(3c)}_\sigma$ should also be understood as expanded to linear order in the argument (keeping higher-order terms is beyond the accuracy of the Fokker-Planck approximation):
\begin{equation}
\bar{\cal P}^{(3a)}_\sigma(1,1')\to {1\over 4}\eta_a\mu_a\,L\!\sum_{232'3'}\!\!^d\,W_\sigma^{(3a)}(1',2',3'|1,2,3)\,\delta_3(\ldots)(k_2-k_{2'})\partial_{k_2}g_\sigma(2)~,\,\,{\rm etc.},
\label{3.23c}
\end{equation}
where the sign $^d$ in $\sum^d$ means that the summation goes over direct processes only, i.e., those with all three transferred momenta $|k_1-k_{1'}|,|k_2-k_{2'}|,|k_3-k_{3'}|\alt 1/a$. The contribution of exchange processes with $|k_2-k_{2'}|\gg 1/a$ to $\bar{\cal P}^{(3a)}_\sigma(1,1')$ in Eq.~(\ref{3.23c}) and to a similar expression for $\bar{\cal P}^{(3b)}_\sigma(1,1')$ is accounted for by the factors $\mu_a=2$ and $\mu_b=2$, respectively. No additional factor on top of $\xi_c$ appears in channel (c), i.e., $\mu_c=1$.  Altogether, one can assign four combinatorial factors $\eta,\xi,\mu,\nu$ to each of the channels as follows:
\begin{equation}\begin{matrix}
  \hspace{1.2cm}& \,\,\,\,\eta\,\,\,\, & \,\,\,\,\xi\,\,\,\, & \,\,\,\,\mu\,\,\,\, & \,\,\,\,\nu\,\,\,\, \vspace{2mm}\\
a & 1/12 & 3   & 2   & 2   \\
b & 1/4  & 1   & 2   & 2   \\
c & 1/2  & 2   & 1   & 1
\end{matrix}\quad ,
\label{3.23d}
\end{equation}
with $\eta\xi\mu\nu=1$. The factors $1/\eta_a=2!3!$, $1/\eta_b=2!2!$, and $1/\eta_c=2!$ give, for a fixed momentum of one electron, the total number of permutations, allowed by exchange scattering, over initial states of two other electrons and final states of all three electrons. The factors $\xi_{a,b,c}$ select the contribution of processes with small momentum transfers $|k_1-k_{1'}|\alt 1/a$ for the diffusing electron (including the exchange processes with large momentum transfers for two other electrons). The factors $\xi_a\mu_a$, $\xi_b\mu_b$, $\xi_c\mu_c$ select the contribution of processes with small momentum transfers for all three electrons $|k_1-k_{1'}|,|k_2-k_{2'}|,|k_3-k_{3'}|\alt 1/a$. The exchange processes thus lead to multiplication of the contribution of the direct processes by factors of 6, 2, 2 in channels (a), (b), (c), respectively. In channels (a) and (b), the factors $\nu_a=1/\eta_a\xi_a\mu_a=2$ and $\nu_b=1/\eta_b\xi_b\mu_b=2$ come from the summation over identical direct scattering processes in which only the labeling is changed as $(2,2')\leftrightarrow (3,3')$ (with exchange processes accounted for by the factors $\xi_{a,b}$ and $\mu_{a,b}$). For $D_\sigma^{(3a),(3b),(3c)}(k_1)$ and for $C_\sigma^{(3a),(3b),(3c)}(k_1)$ with $\bar{\cal P}^{(3a),(3b),(3c)}_\sigma(1,1')$ from Eq.~(\ref{3.23c}) we thus obtain
Eqs.~(\ref{3.83d})-(\ref{3.83c}).

\section{Partial conservation laws}
\label{aC}
\renewcommand{\theequation}{B.\arabic{equation}}
\setcounter{equation}{0}

Apart from the momentum conservation law $\dot{P}_\sigma+\dot{P}_{\bar\sigma}=0$, used in the derivation of Eq.~(\ref{3.16}) for arbitrary $\dot{P}_\sigma$, there are also partial conservation laws for three-particle scattering:
\begin{eqnarray}
&&\dot{P}_\sigma^{(3a)}=0~,
\label{3.70a}\\
&&\dot{P}_\sigma^{(3b)}+\dot{P}_{\bar\sigma}^{(3c)}=0~,
\label{3.70b}
\end{eqnarray}
where $\dot{P}_\sigma^{(3a),(3b),(3c)}$ are the contributions of processes (3a),(3b),(3c) to $\dot{P}_\sigma$, and the conservation law for two-particle scattering, $J_\sigma^{(2)}(k)+J_{\bar\sigma}^{(2)}(k)=0$, which is satisfied ``locally". Equation (\ref{3.70b}) fixes an ``integral" constraint on the terms in the collision integral that describe processes (b) and (c) (a similar constraint on the relation between the two terms is imposed by energy conservation). Microscopically, Eq.~(\ref{3.70b}) relies on the relation between the collision kernels (\ref{3.5b}) and (\ref{3.5c}):
\begin{equation}
W_\sigma^{(3b)}(1',2',3'|1,2,3)=W_{\bar\sigma}^{(3c)}(3',2',1'|3,2,1)~,
\label{3.71}
\end{equation}
valid for arbitrary thermal factors $\zeta_\sigma$, i.e., for arbitrary $\Delta$; which, in turn, relies on the relation between the scattering amplitudes
\begin{equation}
A_{3b,\sigma}^{\rm irr}(1',2',3'|1,2,3)=A_{3c,\bar\sigma}^{\rm irr}(3',2',1'|3,2,1)~.
\label{3.72}
\end{equation}
More specifically, the equation
\begin{eqnarray}
\dot{P}_\sigma^{(3b)}+\dot{P}_{\bar\sigma}^{(3c)}\!\!&=&\!\!-{1\over 4L}\sum_{1231'2'3'}\!k_1\delta_3(\ldots)\left\{\eta_b\,W_\sigma^{(3b)}(1',2',3'|1,2,3)\left[\,g_\sigma(1')+g_{\bar\sigma}(2')+g_{\bar\sigma}(3')-g_\sigma(1)-g_{\bar\sigma}(2)-g_{\bar\sigma}(3)\,
\right]\right.\nonumber\\
&+&\!\!\left.\eta_c\,W_{\bar\sigma}^{(3c)}(1',2',3'|1,2,3)\left[\,g_{\bar\sigma}(1')
+g_{\bar\sigma}(2')+g_\sigma(3')-g_{\bar\sigma}(1)-g_{\bar\sigma}(2)-g_\sigma(3)\,\right]\right\}
\end{eqnarray}
is represented, using Eq.~(\ref{3.71}) and substituting $\eta_b=1/4$ and $\eta_c=1/2$ [Eq.~(\ref{3.23d})], as
\begin{eqnarray}
\dot{P}_\sigma^{(3b)}+\dot{P}_{\bar\sigma}^{(3c)}\!\!&=&\!\!-{1\over 16L}\sum_{1231'2'3'}\!(k_1+2k_3)W_\sigma^{(3b)}(1',2',3'|1,2,3)\delta_3(\ldots)\nonumber\\
&\times&\!\!\left[\,g_\sigma(1')+g_{\bar\sigma}(2')+g_{\bar\sigma}(3')-g_\sigma(1)-g_{\bar\sigma}(2)-g_{\bar\sigma}(3)\,\right]~,
\label{3.73}
\end{eqnarray}
and then, using $W_\sigma^{(3b)}(1',2',3'|1,2,3)=W_\sigma^{(3b)}(1',3',2'|1,3,2)$, as
\begin{eqnarray}
\dot{P}_\sigma^{(3b)}+\dot{P}_{\bar\sigma}^{(3c)}\!\!&=&\!\!-{1\over 16L}\sum_{1231'2'3'}\!(k_1+k_2+k_3)W_\sigma^{(3b)}(1',2',3'|1,2,3)\delta_3(\ldots)\nonumber\\
&\times&\!\!\left[\,g_\sigma(1')+g_{\bar\sigma}(2')+g_{\bar\sigma}(3')-g_\sigma(1)-g_{\bar\sigma}(2)-g_{\bar\sigma}(3)\,\right]~,
\label{3.74}
\end{eqnarray}
which, in turn, using $W_\sigma^{(3b)}(1',2',3'|1,2,3)=W_\sigma^{(3b)}(1,2,3|1',2',3')$, leads to the emergence of the total momentum difference  $k_1+k_2+k_3-k_{1'}-k_{2'}-k_{3'}=0$ as a factor in the integrand, which proves Eq.~(\ref{3.70b}). Equation (\ref{3.70a}) is proven in a similar way.

In the Fokker-Planck limit, Eqs.~(\ref{3.70a}) and (\ref{3.70b}) are also valid exactly.
In particular, $\dot{P}_\sigma^{(3a)}$ is given, in this limit, by
\begin{equation}
\dot{P}_\sigma^{(3a)}=\xi_a\int\!{dk\over 2\pi}\int\!\!\!\!\!\!-\,{dq\over 2\pi}\,q\left[\,{1\over 2}q\,{\cal P}_\sigma^{(3a)}(k,k+q)\partial_kg_\sigma(k)-\bar{\cal P}_\sigma^{(3a)}(k,k+q)\,\right]
\label{3.75}
\end{equation}
with $\bar{\cal P}_\sigma^{(3a)}$ from Eq.~(\ref{3.23c}), i.e., by
\begin{equation}
\dot{P}_\sigma^{(3a)}={1\over 8\nu_aL}\!\sum_{1231'2'3'}\!\!\!\!\!^d\,\,(k_1-k_{1'})W_\sigma^{(3a)}(1',2',3'|1,2,3)\delta_3(\ldots)
\left[\,(k_1-k_{1'})\partial_{k_1}g_\sigma(1)+2(k_2-k_{2'})\partial_{k_2}g_\sigma(2)\,\right]~,
\label{3.76}
\end{equation}
where $\xi_a=3$ and $\nu_a=2$ [Eq.~(\ref{3.23d})].
Using $W^{(3a)}(1',2',3'|1,2,3)=W^{(3a)}(2',1',3'|2,1,3)$, Eq.~(\ref{3.76}) is rewritten as
\begin{equation}
\dot{P}_\sigma^{(3a)}={1\over 8\nu_a L}\!\sum_{1231'2'3'}\!\!\!\!\!^d\,\,(k_1-k_{1'})W_\sigma^{(3a)}(1',2',3'|1,2,3)\delta_3(\ldots)
\left[\,(k_1-k_{1'})+2(k_2-k_{2'})\,\right]\partial_{k_1}g_\sigma(1)~.
\label{3.77}
\end{equation}
Using $W^{(3a)}(1',2',3'|1,2,3)=W^{(3a)}(1',3',2'|1,3,2)$, one of the two terms $k_2-k_{2'}$ in the square brackets in Eq.~(\ref{3.77}) can be changed to $k_3-k_{3'}$; as a result, the total momentum difference $k_1+k_2+k_3-k_{1'}-k_{2'}-k_{3'}=0$ is again obtained as a factor in the integrand. Note that the integral term (\ref{3.23b}) in the Fokker-Planck current in momentum space (\ref{3.21b}) is required for momentum conservation. If it were not for the integral term, $\dot{P}_\sigma^{(3a)}$ would be given by $\int (dk/2\pi)D_\sigma^{(3a)}(k)\partial_k g_\sigma(k)$, which is generically nonzero. In fact, it is the contribution to $\dot{P}_\sigma^{(3a)}$ in Eq.~(\ref{3.77}) that comes from the integral term [which produces the term in the integrand of Eq.~(\ref{3.77}) that is proportional to $(k_1-k_{1'})(k_2-k_{2'})$] that cancels the contribution to $\dot{P}_\sigma^{(3a)}$ of the diffusion term [the corresponding term in the integrand is proportional to $(k_1-k_{1'})^2$].

Similar to Eq.~(\ref{3.76}) for channel (a), the conservation law (\ref{3.70b}), linking channels (b) and (c), is represented in the Fokker-Planck limit as
\begin{eqnarray}
&&\dot{P}_\sigma^{(3b)}+\dot{P}_{\bar\sigma}^{(3c)}={1\over 8L}\sum_{1231'2'3'}\!\!\!\!\!^d\,\,(k_1-k_{1'})\delta_3(\ldots)\nonumber\\
&&\times\left\{{1\over\nu_b}\,W_\sigma^{(3b)}(1',2',3'|1,2,3)\left[\,(k_1-k_{1'})\partial_{k_1}g_\sigma(1)+2(k_2-k_{2'})\partial_{k_2}g_{\bar\sigma}(2)\,\right]
\right.\nonumber\\
&&\!\hspace{2.8mm}+\,{1\over\nu_c}\,W_{\bar\sigma}^{(3c)}(1',2',3'|1,2,3)\left[\,(k_1-k_{1'})
\partial_{k_1}g_{\bar\sigma}(1)+(k_2-k_{2'})\partial_{k_2}g_{\bar\sigma}(2)
+(k_3-k_{3'})\partial_{k_3}g_\sigma(3)\,\right]\Big\}~,
\label{3.78}
\end{eqnarray}
where $\nu_b=2$ and $\nu_c=1$ [Eq.~(\ref{3.23d})].
Substituting Eq.~(\ref{3.71}) in the term describing channel (c), one obtains
\begin{eqnarray}
&&\dot{P}_\sigma^{(3b)}+\dot{P}_{\bar\sigma}^{(3c)}={1\over 16L}\sum_{1231'2'3'}\!\!\!\!\!^d\,\,W_\sigma^{(3b)}(1',2',3'|1,2,3)\delta_3(\ldots)\nonumber\\
&&\times\Big\{(k_1-k_{1'})\left[\,(k_1-k_{1'})\partial_{k_1}g_\sigma(1)+2(k_2-k_{2'})\partial_{k_2}g_{\bar\sigma}(2)\,\right]
\nonumber\\
&&\!\hspace{3mm}+\,2(k_3-k_{3'})\left[\,(k_1-k_{1'})\partial_{k_1}g_\sigma(1)+(k_2-k_{2'})\partial_{k_2}g_{\bar\sigma}(2)
+(k_3-k_{3'})\partial_{k_3}g_{\bar\sigma}(3)\,\right]\Big\}
\label{3.79}
\end{eqnarray}
(the overall factor of 2 in the last line comes from the ratio $\nu_b/\nu_c$). Using $W_\sigma^{(3b)}(1',2',3'|1,2,3)=W_\sigma^{(3b)}(1',3',2'|1,3,2)$, one can change one of two terms $(k_2-k_{2'})\partial_{k_2}g_{\bar\sigma}(2)$ in the first square brackets to $(k_3-k_{3'})\partial_{k_3}g_{\bar\sigma}(3)$ and change the factor $2(k_3-k_{3'})$ in front of the second square brackets to $k_2-k_{2'}+k_3-k_{3'}$. The result is the total momentum difference $k_1+k_2+k_3-k_{1'}-k_{2'}-k_{3'}=0$ emerging as a factor in the integrand.

\section{Limit of $\Delta\to 0$}
\label{aD}
\renewcommand{\theequation}{C.\arabic{equation}}
\setcounter{equation}{0}

In the limit $\Delta\to 0$, the coefficients $a_\sigma$ and $b_\sigma$ from Eqs.~(\ref{3.54a}) and (\ref{3.54b}) do not depend on $\sigma$ and are related by $a_\sigma=-b_\sigma$, so that the determinant of Eq.~(\ref{3.53}), $a_\sigma a_{\bar\sigma}-b_\sigma b_{\bar\sigma}$, vanishes. The solution for $\langle\partial_k{\rm g}_\sigma\rangle$ is not infinite because $c_\sigma=-c_{\bar\sigma}$ at $\Delta=0$. Specifically, if $\Delta=0$, Eq.~(\ref{3.53}) gives for the average of the difference ${\rm g}_\sigma-{\rm g}_{\bar\sigma}$:
\begin{equation}
\langle\partial_k({\rm g}_\sigma-{\rm g}_{\bar\sigma})\rangle={e\over 2}(E_\sigma-E_{\bar\sigma})\left\langle {(1-\tanh)\zeta^{-2}\over {\cal D}^{(2)}\zeta^2+{\cal D}^{(3)}}\right\rangle \left\langle{{\cal D}^{(2)}\zeta^2+2\mu\over {\cal D}^{(2)}\zeta^2+{\cal D}^{(3)}}\right\rangle^{\!\!-1}
\label{3.47a}
\end{equation}
with ${\cal D}^{(3)}=\lambda + \mu$, where we omitted the sign $\sigma$ in $\langle\ldots\rangle_\sigma$, $\lambda_\sigma$, and $\mu_\sigma$.\cite{dmitriev12} Importantly, Eq.~(\ref{3.114}) for $\Delta=0$ gives identically $\partial_k({\rm g}_\sigma+{\rm g}_{\bar\sigma})=\langle\partial_k({\rm g}_\sigma+{\rm g}_{\bar\sigma})\rangle$. This reflects the fact that Eq.~(\ref{3.114}) has a homogeneous solution for ${\rm g}_\sigma$ of the form $\Lambda_1^{\rm reg} k$ with $\Lambda_1^{\rm reg}$ independent of $k$ (the constant term in ${\rm g}_\sigma$ is not allowed by parity, or the particle number conservation for that matter) and $\sigma$. However, $\Lambda_1^{\rm reg}=0$ for the regular at $\omega\to 0$ solution, so that $\partial_k{\rm g}_\sigma=-\partial_k{\rm g}_{\bar\sigma}$ for $\Delta=0$. Using this property in Eq.~(\ref{3.47a}) and substituting the latter in Eq.~(\ref{3.114}), the solution for $\partial_k{\rm g}_\sigma$ at $\Delta=0$ is written as\cite{remark1}
\begin{equation}
\partial_k{\rm g}_\sigma={e\over 4}\left(E_\sigma-E_{\bar\sigma}\right){{\cal T}_1-(\lambda-\mu)\left({\cal T}_1\langle {\cal T}_2\rangle-\langle{\cal T}_1\rangle{\cal T}_2\right)\over 1-(\lambda-\mu)\langle{\cal T}_2\rangle}~,
\label{3.46a}
\end{equation}
where ${\cal T}_1=(1-\tanh){\cal T}_2/\zeta^2$ and ${\cal T}_2=1/[\,{\cal D}^{(2)}\zeta^2+{\cal D}^{(3)}]$.

\end{widetext}

\section{Characteristic momenta for ${\cal C}_\sigma$}
\label{aG}
\renewcommand{\theequation}{D.\arabic{equation}}
\setcounter{equation}{0}

The functions ${\cal D}_\sigma^{(3a),(3b),(3c)}(k_1)$ on the one hand and ${\cal C}_\sigma^{(3a),(3b),(3c)}(k_1)$ on the other are generically determined by different combinations of the characteristic scales of $k_2$ and $k_3$. Namely, ${\cal D}_\sigma^{(3a),(3b),(3c)}(k_1)$ are determined by $k_2$ and $k_3$ close to the Fermi surface(s), but the same is not generically true for ${\cal C}_\sigma^{(3a),(3b),(3c)}(k_1)$. What $k_2$ and $k_3$ give the main contribution to ${\cal C}_\sigma^{(3a),(3b),(3c)}(k_1)$ is determined by the $k$ dependence of the product $\zeta^2_\sigma\partial_k{\rm g}_\sigma$. This dependence changes dramatically with varying strength of intrawire equilibration resulting from three-particle scattering. In the limit of weak equilibration, the product $\zeta^2_\sigma\partial_k{\rm g}_\sigma$ is sharply peaked at the bottom(s) of the spectrum, whereas in the limit of strong equilibration, it is sharply peaked on the Fermi surface(s) [see Eq.~(\ref{3.104}) below]. This simple picture is a hallmark of the clear distinction between equilibration in the stationary or moving (with the drift velocity) frame.\cite{dmitriev12} However, in the extended crossover between these two limiting cases, $\partial_k{\rm g}_\sigma$ shows rather complex behavior. The purpose of Appendix~\ref{aG} is to concisely describe this extended crossover. For clarity, and since the essential features of the transfer of the main weight of $\zeta^2_\sigma\partial_k{\rm g}_\sigma$ from the bottom(s) of the spectrum to the Fermi surface(s) are similar for arbitrary $\Delta$, we describe the steps in the extended crossover for $\Delta=0$ (following the framework developed in Ref.~\onlinecite{dmitriev12}). The most important difference brought about by nonzero $\Delta$ is the appearance of a $k$ independent term in $\zeta^2_\sigma\partial_k{\rm g}_\sigma$ in regime I below, proportional to $\Delta/{\cal D}^{(3)}$.

There are six different regimes, depending on the strength of three-particle equilibration, for the behavior of $\zeta^2\partial_k{\rm g}_\sigma$ (we drop the sign $\sigma$ in $\zeta_\sigma$) as a function of $\epsilon$ for $\epsilon<\epsilon_F$. In the limit of weak three-particle scattering, $\zeta^2\partial_k{\rm g}_\sigma$ decreases as $e^{-\epsilon/T}$ with increasing $\epsilon$ up to $\epsilon_F$:
\begin{eqnarray}
({\rm I})\quad &&\zeta^2\partial_k{\rm g}_\sigma\propto e^{-\epsilon/T}~,\quad \epsilon<\epsilon_F~,
\label{3.94}\\
\nonumber\\
&&{\cal D}^{(3)}e^{\epsilon_F/T}\,,\,{\cal D}_i e^{\epsilon_F/T}(\epsilon_F/T)^{1/2}\ll{\cal D}^{(2)}~.\nonumber
\end{eqnarray}
The condition of applicability of Eq.~(\ref{3.94}), with ${\cal D}_i={\cal D}^{(3a)}-{\cal D}^{(3b)}$, is written under the assumption\cite{dmitriev12} that
the contribution to $\langle\partial_k{\rm g}_\sigma\rangle$ of the divergence of the average on the upper limit of the integration over $|k|\gg k_F$ for two-particle scattering (peculiar to the Fokker-Planck approximation) can be neglected. The divergence is cured either by three-particle scattering or by going beyond the Fokker-Planck expansion of the collision integral. In the former case, the additional condition is $\ln [{\cal D}^{(2)}/{\cal D}^{(3)}]\ll e^{2\epsilon_F/T}$. The sign of ${\cal D}_i$ is generically positive, which is assumed below. Let us also assume that $[{\cal D}_i/{\cal D}^{(3)}](\epsilon_F/T)^{1/2}\!\gg\! 1$, which is satisfied unless the distance between the wires is very small. Then, as three-particle scattering becomes stronger, there appears a plateau in the dependence of $\zeta^2\partial_k {\rm g}_\sigma$ on $\epsilon$ right below $\epsilon_F$:
\begin{eqnarray}
\hspace{-5mm}({\rm II})\quad &&\zeta^2\partial_k{\rm g}_\sigma\propto \!\left\{ \begin{array}{ll}e^{-\epsilon/T}~, & \quad \epsilon<\epsilon_{c1}~,\\
{\rm const}(\epsilon)~, & \quad \epsilon_{c1}<\epsilon<\epsilon_F~,
\end{array}\right.\label{3.95}\\
\nonumber\\
&&{\cal D}^{(3)}e^{\epsilon_F/T}\ll{\cal D}^{(2)}\ll {\cal D}_ie^{\epsilon_F/T}(\epsilon_F/T)^{1/2}~,\nonumber
\end{eqnarray}
where
\begin{equation}
\epsilon_{c1}=\epsilon_F-T\ln\left[{{\cal D}_i\over {\cal D}^{(2)}}e^{\epsilon_F/T}\left({\epsilon_F\over T}\right)^{1/2}\right].
\label{3.96}
\end{equation}
The width of the plateau in Eq.~(\ref{3.95}) grows logarithmically with increasing strength of three-particle scattering. The average $\langle\partial_k{\rm g}_\sigma\rangle$ in regimes I and II is determined by $\epsilon\alt T$. For ${\cal D}^{(3)}e^{\epsilon_F/T}\gg {\cal D}^{(2)}$, we have
\begin{eqnarray}
\hspace{-1mm}({\rm III})\quad &&\zeta^2\partial_k{\rm g}_\sigma\propto \!\left\{\begin{array}{ll}{\rm const}(\epsilon)~, & \quad \epsilon<\epsilon_{c2}~, \\
e^{-\epsilon/T}~, & \quad \epsilon_{c2}<\epsilon<\epsilon_{c3}~, \\
{\rm const}(\epsilon)~, & \quad \epsilon_{c3}<\epsilon<\epsilon_F~,
\end{array}\right.\label{3.97}\\
\nonumber\\
&&{\cal D}_i\left[\dfrac{\epsilon_F}{T}\ln\dfrac{{\cal D}^{(3)}e^{\epsilon_F/T}}{{\cal D}^{(2)}}\right]^{1/2}\!\!\ll{\cal D}^{(2)}\ll {\cal D}^{(3)}e^{\epsilon_F/T}, \nonumber
\end{eqnarray}
where
\begin{eqnarray}
&&\hspace{-8mm}\epsilon_{c2}=T\ln {{\cal D}^{(3)}e^{\epsilon_F/T}\over {\cal D}^{(2)}}=\epsilon_F-T\ln{{\cal D}^{(2)}\over {\cal D}^{(3)}}~,
\label{3.98}\\
&&\hspace{-8mm}\epsilon_{c3}=\epsilon_F-{T\over 2}\ln\left\{\left[{{\cal D}_i\over {\cal D}^{(3)}}\right]^2{\epsilon_F\over T}\ln{{\cal D}^{(3)}e^{\epsilon_F/T}\over {\cal D}^{(2)}}\right\}.
\label{3.99}
\end{eqnarray}
In regime III, $\langle\partial_k{\rm g}_\sigma\rangle$ is determined by $\epsilon<\epsilon_{c2}$, with $\epsilon_{c2}\gg T$. Both plateaus in $\zeta^2\partial_k{\rm g}$ in Eq.~(\ref{3.97}), at the bottom of the spectrum and at the Fermi energy, become wider as ${\cal D}^{(3)}/{\cal D}^{(2)}$ is increased [e.g., for ${\cal D}^{(3)}$ mainly determined by intrawire scattering, which corresponds to ${\cal D}_i/{\cal D}^{(3)}\simeq 1$] and eventually they meet [the left condition of applicability of Eq.~(\ref{3.97}) means that $\epsilon_{c3}>\epsilon_{c2}$]. Note that, if one views ${\cal D}^{(2)}$ as an independent variable, $\epsilon_{c2}$ is a much stronger function of ${\cal D}^{(2)}$ than $\epsilon_{c3}$ [the same is true with respect to ${\cal D}^{(3)}$ for ${\cal D}_i/{\cal D}^{(3)}\simeq 1$]. For this reason, the energy at which the step between the plateaus disappears (so that the dependence on $\epsilon$ in the entire interval $0<\epsilon<\epsilon_F$ is leveled off), $\epsilon_*\simeq \epsilon_F-T\ln [{\cal D}_i\epsilon_F/{\cal D}^{(3)}T]$, is close to the Fermi energy: $T\ll\epsilon_F-\epsilon_*\ll\epsilon_F$. For stronger three-particle scattering, the monotonic decay of $\zeta^2\partial_k{\rm g}_\sigma$ with increasing $\epsilon$ in regimes (I), (II), and (III) changes to a monotonic growth and the step reemerges with an opposite sign. Specifically:
\begin{eqnarray}
\hspace{-8mm}({\rm IV})\quad &&\zeta^2\partial_k{\rm g}_\sigma\propto \!\left\{\begin{array}{ll}{\rm const}(\epsilon)~, & \quad \epsilon<\epsilon_{c3}~, \\
e^{\epsilon/T}~, & \quad \epsilon_{c3}<\epsilon<\epsilon_{c2}~, \\
{\rm const}(\epsilon)~, & \quad \epsilon_{c2}<\epsilon<\epsilon_F~,
\end{array}\right.\label{3.100}\\
&&{\cal D}^{(3)}\ll {\cal D}^{(2)}\ll {\cal D}_i\left[\dfrac{\epsilon_F}{T}\ln\dfrac{{\cal D}^{(3)}e^{\epsilon_F/T}}{{\cal D}^{(2)}}\right]^{1/2}~.\nonumber
\end{eqnarray}
In regime IV, the contribution to $\langle\partial_k{\rm g}_\sigma\rangle$ of electrons on the wide and low plateau ($\epsilon<\epsilon_{c3}$) is much larger than that of electrons on the narrow and high plateau at the Fermi energy. For ${\cal D}^{(3)}\gg {\cal D}^{(2)}$, the plateau at the Fermi energy disappears:
\begin{eqnarray}
\hspace{-8mm}({\rm V})\quad &&\zeta^2\partial_k{\rm g}_\sigma\propto \!\left\{\begin{array}{ll}{\rm const}(\epsilon)~, & \quad \epsilon<\epsilon_{c4}~, \\
e^{\epsilon/T}~, & \quad \epsilon_{c4}<\epsilon<\epsilon_F~,
\end{array}\right.\label{3.101}\\
\nonumber\\
&&{\cal D}^{(3)}e^{-\epsilon_F/T}\epsilon_F/T\ll {\cal D}^{(2)}\ll {\cal D}^{(3)}~,\nonumber
\end{eqnarray}
where
\begin{equation}
\epsilon_{c4}=\epsilon_F-T\ln{{\cal D}^{(3)}\epsilon_F\over {\cal D}^{(2)}T}~.
\label{3.102}
\end{equation}
In regime V, $\langle\partial_k{\rm g}_\sigma\rangle$ is determined by $|\epsilon-\epsilon_F|\sim T$; specifically, the contribution to $\langle\partial_k{\rm g}_\sigma\rangle$ of electrons on the plateau is a factor of $[{\cal D}^{(2)}/{\cal D}^{(3)}](\epsilon_{c4}/\epsilon_F)^{1/2}$ smaller than that of electrons in the spike at the Fermi energy. Note that while ${\cal D}^{(3)}$ can be larger than ${\cal D}^{(2)}$ for sufficiently small $|V_{12}(0)|/|V_{11}(0)|$ (large distance between the wires), the difference ${\cal D}^{(3)}-{\cal D}_i\sim {\cal D}^{(2)}(T/\epsilon_F)^3(v_F/Ta)[V_{11}(0)/v_F]^2$ is much smaller than ${\cal D}^{(2)}$ independently of the ratio ${\cal D}^{(3)}/{\cal D}^{(2)}$. In regime V, therefore, one can neglect the difference between ${\cal D}^{(3)}$ and ${\cal D}_i$. The plateau in Eq.~(\ref{3.101}) becomes narrower as ${\cal D}^{(3)}/{\cal D}^{(2)}$ increases and shrinks to zero at ${\cal D}^{(3)}/{\cal D}^{(2)}\sim e^{\epsilon_F/T}T/\epsilon_F$ [hence the left condition of applicability of Eq.~(\ref{3.101})]. For larger ${\cal D}^{(3)}/{\cal D}^{(2)}$, the product $\zeta^2\partial_k{\rm g}_\sigma$ grows exponentially with increasing $\epsilon$ in the whole range of $\epsilon$ between $0$ and $\epsilon_F$:
\begin{eqnarray}
\hspace{-8mm}({\rm VI})\quad &&\zeta^2\partial_k{\rm g}_\sigma\propto e^{\epsilon/T}~,\quad \epsilon<\epsilon_F~,\label{3.103}\\
&&{\cal D}^{(2)}\ll {\cal D}^{(3)}e^{-\epsilon_F/T}\epsilon_F/T~.\nonumber
\end{eqnarray}
Thus, the evolution of the behavior of $\zeta^2\partial_k{\rm g}_\sigma$ as a function of $\epsilon$ between regimes I and VI results in the change of the sign in the argument of the exponential function:
\begin{equation}
e^{-\epsilon/T} \,\,{\rm (I)}\,\,\leftrightarrow \,\,e^{\epsilon/T} \,\,{\rm (VI)}~.
\label{3.104}
\end{equation}

In regimes V and VI [i.e., for ${\cal D}^{(3)}\gg {\cal D}^{(2)}$], the main contribution to $\langle\partial_k{\rm g}_\sigma\rangle$ comes from electrons on the Fermi level, with $|\epsilon-\epsilon_F|\alt T$ and ${\rm g}_\sigma$ being a smooth function of $\epsilon$ on this scale. Therefore, in these regimes, the representation of ${\cal C}_\sigma^{(3a),(3b),(3c)}(k)$ in terms of the products of the diffusion coefficients ${\cal D}_\sigma^{(3a),(3b),(3c)}(k)$ and the averages $\langle\partial_k{\rm g}_\sigma\rangle$ is accurate. Now, the integral term ${\cal C}_\sigma$ is only important in the calculation of $\rho_{\rm D}$ for ${\cal D}^{(3)}\agt {\cal D}^{(2)}e^{\epsilon_F/T}(T/\epsilon_F)^{3/2}$, i.e., deep in regime V and in regime VI. Hence, the use of the representation
(\ref{3.116}) is justified in the calculation of $\rho_{\rm D}$. It is also worth noting that, in regimes I, II, and for $\epsilon_{c2}\ll\epsilon_F$ in regime III, the representation of ${\cal C}_\sigma$ as a linear combination of $\langle\partial_k{\rm g}_\sigma\rangle$ and $\langle\partial_k{\rm g}_{\bar\sigma}\rangle$ is parametrically accurate as well; however, the prefactors of $\langle\partial_k{\rm g}_\sigma\rangle$ and $\langle\partial_k{\rm g}_{\bar\sigma}\rangle$ in this combination are not given by the diffusion coefficients ${\cal D}^{(3a),(3b),(3c)}_\sigma$, in contrast to the model of Sec.~\ref{s3}. This is because the model neglects the dependence of ${\cal D}^{(3)}_\sigma$ on $k$.

\vspace{4mm}

\end{appendix}

\clearpage

\begin{widetext}

\begin{center}
ONLINE SUPPLEMENTARY MATERIAL

\vspace{6mm}

{\bf Ultranarrow resonance in Coulomb drag between quantum wires\\
at coinciding densities}

\vspace{6mm}

A.P. Dmitriev, I.V. Gornyi, and D.G. Polyakov

\vspace{4mm}
\end{center}

\renewcommand{\thepage}{S\arabic{page}}
\renewcommand{\theequation}{S\arabic{equation}}
\renewcommand{\thefigure}{S\arabic{figure}}

\setcounter{page}{1}
\setcounter{section}{0}
\setcounter{equation}{0}
\setcounter{figure}{0}

Supplementary Material gives some technical details relevant to the model from Sec.~\ref{s7}. Specifically, the exact functions ${\cal D}^{(3)}_\sigma(k)$ and ${\cal C}_\sigma(k)$ are represented in the form that shows that these are slow functions of $k$ compared to the exponentials $\zeta^2_\sigma(k)$. The three-particle scattering amplitudes, which enter ${\cal D}^{(3)}_\sigma(k)$ and ${\cal C}_\sigma(k)$, are written in a compact form in the limit of small momentum transfer.


{\it Partial contributions to ${\cal D}^{(3)}_\sigma$ and ${\cal C}_\sigma$.} For given three initial momenta $k_1,k_2,k_3$ and the momentum transfer $q=k_{1'}-k_1$, there are two solutions, allowed by conservation of energy and momentum, for the momentum transfer $q_2=k_{2'}-k_2$:
\begin{equation}
q_2(q,k_1,k_2,k_3)=-{1\over 2}(k_2-k_3+q)\pm \left[\,{1\over 4}(k_2-k_3+q)^2-q(k_1-k_3+q)\,\right]^{1/2}~.
\label{3.80}
\end{equation}
For given $k_1,k_2,k_3$, there are thus two branches of $q_2$ as a function of $q$, depending on the sign of the square root in Eq.~(\ref{3.80}). The two branches terminate at two common points---one point with $q>0$, the other with $q<0$---at which the argument of the square root becomes zero.
For $q\to 0$, either $q_2\to 0$ (branch 1) or $q_2\to k_3-k_2$ (branch 2). Specifically, for $|q(k_1-k_3+q)|\ll (k_2-k_3+q)^2$, we have either $q_2\simeq -q(k_1-k_3+q)/(k_2-k_3+q)$ on branch 1 or $q_2\simeq k_3-k_2-q$ on branch 2. For branch 1, the relation between $q_2$ and $q$ in the limit $q\to 0$ reduces to
\begin{equation}
q_2\simeq -q\,{k_1-k_3\over k_2-k_3}~,\qquad |q|\ll |k_2-k_3|\,,\,|k_1-k_3|\,,\,(k_2-k_3)^2/|k_1-k_3|~.
\label{3.81}
\end{equation}
In the limit of $k_1-k_3\to 0$ taken before the limit $q\to 0$, the relation between $q_2$ and $q$ changes from linear in Eq.~(\ref{3.81}) to quadratic:
\begin{equation}
q_2\simeq -{q^2\over k_2-k_3}~,\qquad |k_1-k_3|\ll q\ll |k_2-k_3|~.
\label{3.81a}
\end{equation}
Similar relations between $q_3$ and $q$ are given by Eqs.~(\ref{3.81}) and (\ref{3.81a}) with the change $k_2\leftrightarrow k_3$.

Scattering on branch 1 gives the main contribution to $\rho_{\rm D}$ for $T\ll\epsilon_F$ in the limits of small and large distances between the wires (see Sec.~IIID in Ref.~\onlinecite{dmitriev12} for more detail), with scattering on branch 2 [specifically, in channel (c)] being only relevant in the crossover between the limits (where it does not lead to any qualitatively important changes). Below, $D_{\sigma 1}^{(3a),(3b),(3c)}$ and $C_{\sigma 1}^{(3a),(3b),(3c)}$ denote the contributions to $D_\sigma^{(3a),(3b),(3c)}$ and $C_\sigma^{(3a),(3b),(3c)}$ of scattering on branch 1. In the limit $q\to 0$, the delta function $\delta_3(\ldots)$ for $q_2$ on branch 1 is written as
\begin{equation}
\delta_3(\ldots)\to {m\over |k_2-k_3|}\delta\left(q_2+q\,{k_1-k_3\over k_2-k_3}\right)={m\over |k_2-k_3|}\delta\left(q_3-q\,{k_1-k_2\over k_2-k_3}\right)~.
\label{3.83}
\end{equation}
We have, then, in the limit $q\to 0$:
\begin{eqnarray}
&&D_{\sigma 1}^{(3a),(3b),(3c)}(k_1)={m\over 8\nu_{a,b,c}}\sum_{231'2'3'}\,\!\!\!\!\!^d \,W_\sigma^{(3a),(3b),(3c)}(1',2',3'|1,2,3)\,{q^2\over |k_2-k_3|}\,\delta\left(q_2+q\,{k_1-k_3\over k_2-k_3}\right)~,
\label{3.84d}\\
&&C_{\sigma 1}^{(3a)}(k_1)={m\over 4\nu_a}\sum_{231'2'3'}\,\!\!\!\!\!^d \,W_\sigma^{(3a)}(1',2',3'|1,2,3)\,q^2\,\delta\left(q_2+q\,{k_1-k_3\over k_2-k_3}\right)
{{\rm sgn}(k_2-k_3)\over (k_2-k_3)^2}\,(k_1-k_3)\,\partial_{k_2}g_\sigma(2)~,
\label{3.84}\\
&&C_{\sigma 1}^{(3b)}(k_1)={m\over 4\nu_b}\sum_{231'2'3'}\,\!\!\!\!\!^d \,W_\sigma^{(3b)}(1',2',3'|1,2,3)\,q^2\,\delta\left(q_2+q\,{k_1-k_3\over k_2-k_3}\right){{\rm sgn}(k_2-k_3)\over (k_2-k_3)^2}\,(k_1-k_3)\,\partial_{k_2}g_{\bar\sigma}(2)~,
\label{3.84a}\\
&&C_{\sigma 1}^{(3c)}(k_1)={m\over 8\nu_c}\sum_{231'2'3'}\,\!\!\!\!\!^d \,W_\sigma^{(3c)}(1',2',3'|1,2,3)\,q^2\,\delta\left(q_2+q\,{k_1-k_3\over k_2-k_3}\right)\nonumber\\
&&\hspace{6.5cm}\times\,{{\rm sgn}(k_2-k_3)\over (k_2-k_3)^2}\left[\,(k_1-k_3)\,\partial_{k_2}g_\sigma(2)-(k_1-k_2)\,\partial_{k_3}g_{\bar\sigma}(3)\,\right]~.
\label{3.84b}
\end{eqnarray}
Reducing the number of summations over momenta to three, and taking the limit $q\to 0$ also in the thermal functions $\zeta_\sigma(1')\to\zeta_\sigma(1)$, etc., we have for ${\cal D}_{\sigma 1}^{(3a),(3b),(3c)}(k_1)$ [Eq.~(\ref{3.40b})]:
\begin{eqnarray}
{\cal D}_{\sigma 1}^{(3a)}(k_1)&=&{1\over 128}\left({m\over 2\pi}\right)^3\int\!dk_2\,\zeta^2_\sigma(2)\int\!dk_3\,\zeta^2_\sigma(3){1\over |k_2-k_3|}
d_\sigma^{a}(k_1,k_2,k_3)~,\label{3.107a}\\
{\cal D}_{\sigma 1}^{(3b)}(k_1)&=&{1\over 128}\left({m\over 2\pi}\right)^3\int\!dk_2\,\zeta^2_{\bar\sigma}(2)\int\!dk_3\,\zeta^2_{\bar\sigma}(3){1\over |k_2-k_3|}d_\sigma^{b}(k_1,k_2,k_3)~,\label{3.107b}\\
{\cal D}_{\sigma 1}^{(3c)}(k_1)&=&{1\over 64}\left({m\over 2\pi}\right)^3\int\!dk_2\,\zeta^2_\sigma(2)\int\!dk_3\,\zeta^2_{\bar\sigma}(3){1\over |k_2-k_3|}d_\sigma^{c}(k_1,k_2,k_3)~,
\label{3.107c}
\end{eqnarray}
where the ``momentum-resolved" diffusion coefficients $d^{a,b,c}_\sigma(k_1,k_2,k_3)$ are given by
\begin{equation}
d^{a,b,c}_\sigma(k_1,k_2,k_3)={1\over 2}\int\!dq\,q^2|{\cal A}^{a,b,c}_{\sigma}(q,k_1,k_2,k_3)|^2~.
\label{3.105d}
\end{equation}
The function ${\cal A}^{a}_{\sigma}(q,k_1,k_2,k_3)$ in Eq.~(\ref{3.105d}) is defined [those for channels (b) and (c) are defined similarly] by
\begin{equation}
\left(A^{\rm irr}_{3a,\sigma}\right)_{\rm dir}={m\over L^2}\,\delta_K{\cal A}^{a}_{\sigma}~,
\label{3.106}
\end{equation}
where the Kronecker symbol $\delta_K=\delta_{k_1+k_2+k_3,k_{1'}+k_{2'}+k_{3'}}$, with six momenta in the amplitude $\left(A^{\rm irr}_{3a,\sigma}\right)_{\rm dir}$ taken on the mass-shell at $q+q_2+q_3=0$, which fixes $q_2$ and $q_3$ as functions of $q$, $k_1$, $k_2$, and $k_3$; hence four arguments of ${\cal A}^{a}_{\sigma}$. The integral terms ${\cal C}_{\sigma 1}^{(3a),(3b),(3c)}(k_1)$ [Eq.~(\ref{3.40c})] are expressed through $d^{a,b,c}_\sigma(k_1,k_2,k_3)$ as follows:
\begin{eqnarray}
{\cal C}_{\sigma 1}^{(3a)}(k_1)&=&{1\over 64}\left({m\over 2\pi}\right)^3\int\!dk_2\,\zeta_\sigma^2(2)\,\partial_{k_2}g_\sigma(2)F^{a}_{\sigma}(k_1,k_2)~,
\label{3.104a}\\
{\cal C}_{\sigma 1}^{(3b)}(k_1)&=&{1\over 64}\left({m\over 2\pi}\right)^3\int\!dk_2\,\zeta_{\bar\sigma}^2(2)\,\partial_{k_2}g_{\bar\sigma}(2)F^{b}_{\sigma}(k_1,k_2)~,
\label{3.104b}\\
{\cal C}_{\sigma 1}^{(3c)}(k_1)&=&{1\over 64}\left({m\over 2\pi}\right)^3 \left[\int\!dk_2\,\zeta_\sigma^2(2)\,\partial_{k_2}g_\sigma(2)F^{c}_{\sigma}(k_1,k_2)-\int\!dk_3\,\zeta_{\bar\sigma}^2(3)\,
\partial_{k_3}g_{\bar\sigma}(3)\bar{F}^{c}_{\sigma}(k_1,k_3)\right]~,
\label{3.104c}
\end{eqnarray}
where
\begin{eqnarray}
&&F_{\sigma}^{a}(k_1,k_2)=\int\!dk_3\,\zeta^2_\sigma(3){{\rm sgn}(k_2-k_3)\over (k_2-k_3)^2}\,(k_1-k_3)\,d^{a}_\sigma(k_1,k_2,k_3)~,
\label{3.105a}\\
&&F_{\sigma}^{b}(k_1,k_2)=\int\!dk_3\,\zeta^2_{\bar\sigma}(3){{\rm sgn}(k_2-k_3)\over (k_2-k_3)^2}\,(k_1-k_3)\,d^{b}_\sigma(k_1,k_2,k_3)~,
\label{3.105b}\\
&&F_{\sigma}^{c}(k_1,k_2)=\int\!dk_3\,\zeta^2_{\bar\sigma}(3){{\rm sgn}(k_2-k_3)\over (k_2-k_3)^2}\,(k_1-k_3)\,d^{c}_\sigma(k_1,k_2,k_3)~,
\label{3.105c1}\\
&&\bar{F}_{\sigma}^{c}(k_1,k_3)=\int\!dk_2\,\zeta^2_\sigma(2){{\rm sgn}(k_2-k_3)\over (k_2-k_3)^2}\,(k_1-k_2)\,d^{c}_\sigma(k_1,k_2,k_3)~.
\label{3.105c2}
\end{eqnarray}


{\it Three-particle scattering amplitudes for soft collisions.} The amplitude of direct three-particle scattering\cite{lunde07} in channel (a) can be represented on the mass-shell as
\begin{equation}
\left(A_{3a,\sigma}^{\rm irr}\right)_{\rm dir}={m\over L^2}\,\delta_K\!
\left[\,{V_{\sigma\sigma}(q)V_{\sigma\sigma}(q_3)\over (k_{3'}-k_2)(k_{1'}-k_2)}+{V_{\sigma\sigma}(q)V_{\sigma\sigma}(q_2)\over (k_{2'}-k_3)(k_{1'}-k_3)}+{V_{\sigma\sigma}(q_2)V_{\sigma\sigma}(q_3)\over (k_{3'}-k_1)(k_{2'}-k_1)}\,\right].
\label{3.85}
\end{equation}
[this representation is obtainable by using the mass-shell relations given by Eq.~(3.36) in Ref.~\onlinecite{dmitriev12}].
Taking $\left(A_{3a,\sigma}^{\rm irr}\right)_{\rm dir}$ on branch 1 of the mass-shell and neglecting transferred momenta (omitting the sign $'$ in $k_{1'},k_{2'},k_{3'}$) in the denominators of the partial contributions to $\left(A_{3a,\sigma}^{\rm irr}\right)_{\rm dir}$, we have the scattering amplitude for small $q=k_{1'}-k_1$:
\begin{equation}
\left(A_{3a,\sigma}^{\rm irr}\right)_{\rm dir}\to {m\over L^2}\,\delta_K{1\over (k_2-k_3)^2}
\left[\,-{V_{\sigma\sigma}(q)V_{\sigma\sigma}(q\chi)\over \chi}+{V_{\sigma\sigma}(q)V_{\sigma\sigma}(q+q\chi)\over 1+\chi}+{V_{\sigma\sigma}(q\chi)V_{\sigma\sigma}(q+q\chi)\over \chi (1+\chi)}\,\right],
\label{3.86}
\end{equation}
where
\begin{equation}
\chi={k_1-k_2\over k_2-k_3}~.
\label{3.87}
\end{equation}
Similar in channels (b) and (c):
\begin{eqnarray}
\left(A_{3b,\sigma}^{\rm irr}\right)_{\rm dir}\!\!&\to&\!\! {m\over L^2}\,\delta_K{1\over (k_2-k_3)^2}\left[\,-{V_{\sigma\bar\sigma}(q)V_{\bar\sigma\bar\sigma}(q\chi)\over \chi}+{V_{\sigma\bar\sigma}(q)V_{\bar\sigma\bar\sigma}(q+q\chi)\over 1+\chi}+{V_{\sigma\bar\sigma}(q\chi)V_{\sigma\bar\sigma}(q+q\chi)\over \chi (1+\chi)}\,\right],\label{3.90}\\
\left(A_{3c,\sigma}^{\rm irr}\right)_{\rm dir}\!\!&\to&\!\!
{m\over L^2}\,\delta_K{1\over (k_2-k_3)^2}\left[\,-{V_{\sigma\sigma}(q)V_{\sigma\bar\sigma}(q\chi)\over \chi}+{V_{\sigma\bar\sigma}(q)V_{\sigma\bar\sigma}(q+q\chi)\over 1+\chi}+{V_{\sigma\bar\sigma}(q\chi)V_{\sigma\sigma}(q+q\chi)\over \chi (1+\chi)}\,\right].
\label{3.91}
\end{eqnarray}
For $|k_1|\ll |k_2|\simeq |k_3|$ with $k_2\simeq -k_3$ ($\chi\simeq -1/2$), Eq.~(\ref{3.86}) gives
\begin{equation}
\left(A_{3a,\sigma}^{\rm irr}\right)_{\rm dir}\to {m\over L^2}\delta_K{1\over k_2^2}\,V_{\sigma\sigma}(q/2)\left[\,V_{\sigma\sigma}(q)-V_{\sigma\sigma}(q/2)\,\right]~,\qquad
|k_1|\ll |k_2|\simeq |k_3|~,\,\,\,\,k_2\simeq -k_3~,
\label{3.88}
\end{equation}
which coincides with the result of Ref.~\onlinecite{dmitriev12} for $k_2$ on the Fermi surface. For $|k_1-k_2|\ll |k_2-k_3|$ ($\chi\to 0$), Eq.~(\ref{3.86}) reduces to
\begin{equation}
\left(A_{3a,\sigma}^{\rm irr}\right)_{\rm dir}\to {m\over L^2}\delta_K{1\over (k_2-k_3)^2}\left\{\,qV_{\sigma\sigma}(0){dV_{\sigma\sigma}(q)\over dq}-V_{\sigma\sigma}(q)\left[\,V_{\sigma\sigma}(0)-V_{\sigma\sigma}(q)\,\right]\right\}~,\qquad |k_1-k_2|\ll |k_2-k_3|~.
\label{3.89}
\end{equation}
Note that the singularities in Eq.~(\ref{3.86}) at $k_1\to k_2$ cancel out in Eq.~(\ref{3.89}), although the singularities at $k_1\to k_{2'}$ and $k_{1'}\to k_2$ in the exact expression for the direct amplitude in Eq.~(\ref{3.85}) do not. Recall, however, that the order of taking limits in the derivation of Eq.~(\ref{3.86}) ($q\to 0$ before any other limit) implies that $|k_1-k_2|\gg |q|$ in Eq.~(\ref{3.89}). The singularities at $k_1-k_2\to -q$ and $q_2$ are, in fact, cancelled by exchange processes. Moreover, in all channels (a), (b), and (c), the singularities are regularized in the collision integral as described in the end of Sec.~\ref{s3}.\cite{resibois65,bezzerides68,dmitriev12}
Therefore, in the Fokker-Planck limit, for the calculation of the contribution to $C_{\sigma 1}^{(3a)}(k_1)$ of scattering with $|k_1-k_2|\ll |k_2-k_3|$, it is legitimate to use Eq.~(\ref{3.89}) for $k_1$ arbitrarily close to $k_2$.

In channels (b) and (c), for $|k_1|\ll |k_2|\simeq |k_3|$ with $k_2\simeq -k_3$, Eqs.~(\ref{3.90}) and (\ref{3.91}) give
\begin{eqnarray}
&&\left(A_{3b,\sigma}^{\rm irr}\right)_{\rm dir}\to {m\over L^2}\delta_K{1\over k_2^2}\left[\, V_{\bar\sigma\bar\sigma}(q/2)V_{\sigma\bar\sigma}(q)-V_{\sigma\bar\sigma}^2(q/2)\,\right]~,\nonumber\\
&&\left(A_{3c,\sigma}^{\rm irr}\right)_{\rm dir}\to {m\over L^2}\delta_K{1\over 2k_2^2}\,V_{\sigma\bar\sigma}(q/2)\left[\,V_{\sigma\sigma}(q)+V_{\sigma\bar\sigma}(q)-2V_{\sigma\sigma}(q/2)\,\right]~,\qquad
|k_1|\ll |k_2|\simeq |k_3|~,\,\,\,\,k_2\simeq -k_3~,
\label{3.88a}
\end{eqnarray}
in accordance with Ref.~\onlinecite{dmitriev12}. For $|k_1-k_2|\ll |k_2-k_3|$, we have:
\begin{eqnarray}
&&\left(A_{3b,\sigma}^{\rm irr}\right)_{\rm dir}\to {m\over L^2}\delta_K{1\over (k_2-k_3)^2}\left\{\,-{V_{\sigma\bar\sigma}(q)\left[\,V_{\bar\sigma\bar\sigma}(0)-V_{\sigma\bar\sigma}(0)\,\right]\over \chi}+qV_{\sigma\bar\sigma}(0){dV_{\sigma\bar\sigma}(q)\over dq}-V_{\sigma\bar\sigma}(q)\left[\,V_{\sigma\bar\sigma}(0)-V_{\bar\sigma\bar\sigma}(q)\,\right]\right\}~,
\nonumber\\
&&\left(A_{3c,\sigma}^{\rm irr}\right)_{\rm dir}\to {m\over L^2}\delta_K{1\over (k_2-k_3)^2}\left[\,qV_{\sigma\bar\sigma}(0){dV_{\sigma\sigma}(q)\over dq}-V_{\sigma\sigma}(q)V_{\sigma\bar\sigma}(0)+V_{\sigma\bar\sigma}^2(q)\,\right]~,\label{3.93} \qquad |k_1-k_2|\ll |k_2-k_3|~.
\label{3.92}
\end{eqnarray}
The singularities at $k_1\to k_2$ in the direct amplitude cancel out in channel (c), similar to channel (a), but do not in channel (b).

\end{widetext}


\begin{thebibliography}{99}
\bibitem{remark6} More accurately, the ansatz $g_1(k)-g_2(k)={\rm C}$, where $\rm C$ is an arbitrary constant of $k$, also leads to the vanishing of ${\rm st}_\sigma^{(2)}$ from Eq.~(\ref{4.10a}). In fact, ${\rm C}\neq 0$ would mean a nonequilibrium (proportional to the current) splitting of the chemical potentials in two wires, additional to $\Delta$. In the thermodynamic limit, however, a flow of electrons through the wire cannot produce ${\rm C}\neq 0$, because otherwise it would be accompanied by a change of the electron density, which is fixed in each of the wires.
\bibitem{remark7} For arbitrary $T/\epsilon_{F\sigma}$, the product $\bar{n}{\cal I}(\Delta/2T)$ in Eq.~(\ref{4.23}), which comes from the singular part of $\sigma_{12}$, should be substituted by the integral $(1/4\pi mT)\!\int\! dk\, k^2\zeta_1^2\zeta_2^2/(\zeta_1^2+\zeta_2^2)$. The $1/\omega^2$ divergency of $\det\hat\sigma$ for two-particle scattering at $n_1\neq n_2$ occurs for arbitrary $T/\epsilon_{F\sigma}$.
\bibitem{remark5} In fact, Ref.~\onlinecite{fuchs05} calculated the drag resistance at $\Delta\neq 0$ for a finite system perturbatively (to the lowest order) in the strength of backscattering and associated the drag resistivity with the drag resistance per unit length. However, even if the drag resistance of a small system is a linear function of the system size, the perturbative calculation of this kind does not necessarily describe bulk drag, i.e., does not give the correct result for the drag resistivity (defined conventionally, similar to the diagonal resistivity, through the response functions of a macroscopically homogeneous infinite system, with the system size being larger than any relevant relaxation length).
\bibitem{remark2} If one proceeds with the representation (\ref{3.116}) in terms of four constants characterizing three-particle scattering $\lambda_\sigma$ and $\mu_\sigma$, then Eq.~(\ref{3.53}) allows for a regular solution for any $\Delta\neq 0$ but generically shows a singularity in the solution at $\Delta=0$ of the $1/\Delta$ type. That is, the finite solution\cite{dmitriev12} at $\Delta=0$ appears to be squeezed between the two divergencies at $\Delta\to\pm 0$.
\bibitem{remark1} The Fokker-Planck solution for ${\rm g}_\sigma(k)$ from Eq.~(\ref{3.46a}) for ${\cal D}^{(3)}\neq 0$ diverges in the limit $|k|\to\infty$ as a linear function of $k$, whereas it diverges as $e^{k^2/2mT}/k$ if only two-particle scattering is present. The (linear in the upper limit of the integration over $k$) divergence of $\langle\partial_k{\rm g}_\sigma\rangle$ in the latter case is an artifact of the Fokker-Planck approximation\cite{dmitriev12} (it is cut off by going beyond this approximation for $|k|\agt mTa$).
\bibitem{remark4} The drift regime for Coulomb drag between quantum wires was proposed in Ref.~\onlinecite{pustilnik03} within the framework of the conventional (``perturbative") theory of Coulomb drag. As mentioned in Sec.~\ref{intro}, the applicability of the perturbative theory in one dimension is, however, severely restricted by slow thermalization between electrons [as the condition above Eq.~(\ref{3.184}) shows] and corresponds to the limit of a large separation between the wires. See Ref.~\onlinecite{dmitriev12} for a detailed discussion of the inherent relation between thermalization and Coulomb drag.
\end{thebibliography}
\end{document}